\documentclass[journal]{IEEEtran}
\usepackage[utf8]{inputenc} 
\usepackage[T1]{fontenc}
\usepackage{url}
\usepackage{ifthen}
\usepackage{cite}
\usepackage{amssymb}
\usepackage{amsfonts}
\usepackage{amsthm}
\usepackage[cmex10]{amsmath}
\usepackage{dsfont}
\usepackage{balance}
\usepackage{xfp}

\usepackage{graphicx}
\usepackage{hyperref}                                   
\hypersetup{colorlinks=true,allcolors=blue}
\usepackage{xargs}                                      
\usepackage{xcolor}                                     
\usepackage{algorithm}                                  
\usepackage{algpseudocode}                              
\usepackage{authblk}                                    
\usepackage[printonlyused]{acronym}                     
\usepackage{enumitem}                                   
\usepackage{lipsum}                                     
\usepackage{verbatim}
\usepackage{xspace}

\newtheorem{theorem}{Theorem}
\newtheorem{corollary}{Corollary}

\newtheorem{lemma}{Lemma}

\newtheorem{proposition}{Proposition}

\acrodef{HT}[HT]{Hypothesis Testing}
\acrodef{SHT}[SHT]{Sequential \ac{HT}}
\acrodef{KLD}[KLD]{Kullback-Leibler Divergence}
\acrodef{DM}[DM]{Decision-Maker}
\acrodef{PDF}[PDF]{Probability Density Function}
\acrodef{PMF}[PMF]{Probability Mass Function}
\acrodef{CDF}[CDF]{Cumulative Distribution Function}
\acrodef{CCDF}[CCDF]{Complementary \ac{CDF}}
\acrodef{LRT}[LRT]{Likelihood Ratio Test}
\acrodef{LLR}[LLR]{Log-Likelihood Ratio}
\acrodef{LLRT}[LLRT]{\ac{LLR} Test}
\acrodef{SPRT}[SPRT]{Sequential Probability Ratio Test}
\acrodef{MSPRT}[MSPRT]{Multihypothesis \ac{SPRT}}
\acrodef{MAP}[MAP]{Maximum A Posteriori}
\acrodef{ML}[ML]{Maximum Likelihood}
\acrodef{AI}[AI]{Artificial Intelligence}
\acrodef{ML}[ML]{Machine Learning}
\acrodef{ABR}[ABR]{Average Bayes Risk}
\acrodef{AEP}[AEP]{Asymptotic Equipartition Property}
\acrodef{MWDT}[MWDT]{Minimal Weight Decision Tree}
\acrodef{TVD}[TVD]{Total Variation Distance}
\acrodef{ISIT}[ISIT]{IEEE International Symposium on Information Theory}
\acrodef{PHI}[PHI]{Pruning Hypotheses Iteratively}
\acrodef{DELTA}[DELTA]{Distribution-based Early Labeling for Tapered Acquisitions}
\acrodef{IOTA}[IOTA]{Independent One-at-a-Time Action}
\acrodef{SEF}[SEF]{Scalar Exponential Family}
\acrodef{MGF}[MGF]{Moment-Generating Function}
\acrodef{SNR}[SNR]{Signal-to-Noise Ratio}

\newcommand{\PHIDELTA}{$\Phi$-$\Delta$\xspace}

\def\HT{\ac{HT}\xspace}
\def\SHT{\ac{SHT}\xspace}
\def\KLDtxt{\ac{KLD}\xspace}
\def\KLDstxt{\acp{KLD}\xspace}
\def\DM{\ac{DM}\xspace}
\def\PDF{\ac{PDF}\xspace}

\def\CDF{\ac{CDF}\xspace}

\def\LLR{\ac{LLR}\xspace}
\def\LLRs{\acp{LLR}\xspace}

\def\SPRT{\ac{SPRT}\xspace}
\def\MSPRT{\ac{MSPRT}\xspace}
\def\ML{\ac{ML}\xspace}
\def\AI{\ac{AI}\xspace}
\def\ABR{\ac{ABR}\xspace}

\def\MWDT{\ac{MWDT}\xspace}
\def\TVDtxt{\ac{TVD}\xspace}

\def\MSPRT{\ac{MSPRT}\xspace}
\def\PHI{\ac{PHI}\xspace}

\def\IOTA{\ac{IOTA}\xspace}
\def\SEF{\ac{SEF}\xspace}
\def\MGF{\ac{MGF}\xspace}
\def\SNR{\ac{SNR}\xspace}

\newif\ifBlind
\Blindtrue
\Blindfalse 

\newif\ifShowProofSkech
\ShowProofSkechtrue

\newif\ifPaperAward
\PaperAwardtrue
\PaperAwardfalse 

\newif\ifShowAppendix
\ShowAppendixtrue

\newif\ifReferToAppendix
\ReferToAppendixtrue

\newif\ifShowSupp
\ShowSupptrue

\newcommandx{\myVec}[2][2=]{{\underline{#1}_{#2}}}                              
\newcommandx{\myMat}[2][2=]{{\mathbf{#1}_{#2}}}                                 
\newcommandx{\vectorComponent}[4][3=, 4=]{{[\myVec{#1}_{#3}]_{#2}^{#4}}}        
\newcommandx{\matrixComponent}[4][4=]{{[\myMat{#1}_{#4}]_{#2,#3}}}              
\newcommand{\KLD}[2]{\mathcal{D}_{KL} ( #1 \Vert #2 )}                          

\newcommand{\KLDthetai}[0]{\KLD{f_\theta^{a}}{f_i^{a}}}
\newcommand{\KLDthetaj}[0]{\KLD{f_\theta^{a}}{f_j^{a}}}

\newcommand{\DKLD}[4]{\Delta\mathcal{D}_{#2#3|#1}(#4)}

\newcommand{\intSet}[2]{\{#1, \inteval{#1 + 1}, \dots, #2\}}                    
\newcommand{\expVal}[1]{\mathbb{E}\left[#1\right]}                              
\newcommand{\expValDist}[2]{\mathbb{E}_{#2}\left[#1\right]}
\newcommand{\indicator}[1]{ \mathds{1}{\left\{#1\right\}} }                     
\newcommand{\prob}[1]{\operatorname{\mathbb{P}}\left( #1 \right)}               
\newcommand{\bigO}[1]{\operatorname{\mathcal{O}}\left( #1 \right)}              
\newcommand{\littleO}[1]{\operatorname{o}\left( #1 \right)}                     
\newcommand{\bigTheta}[1]{\operatorname{\Theta}\left( #1 \right)}               

\newcommand{\NaP}[0]{N_a^{\Phi}}

\newcommand{\NaPD}[0]{N_a^{\text{\PHIDELTA}}}

\newcommand{\hII}{H_i}
\newcommand{\hJJ}{H_j}
\newcommand{\hKK}{H_k}
\newcommand{\hTheta}{H_{\theta}}
\newcommand{\hAlive}{H_{\mathrm{alive}}}

\newcommand{\argmax}[2]{\mathop{\operatorname{argmax}}_{#2} #1}
\newcommand{\argmin}[2]{\mathop{\operatorname{argmin}}_{#2} #1}
\newcommand{\abs}[1]{| #1 |}

\newcommand{\equivClass}[3]{\operatorname{class}\left(#1, #2, #3\right)}
\newcommand{\equivCluster}[4]{\mathcal{C}_{#1}^{#3}\left(#2\right)}
\newcommand{\equivClassIA}[1]{\equivClass{i}{#1}{a}}

\newcommand{\equivClusterIAEps}[1]{\equivCluster{i}{#1}{a}{\varepsilon_a}}
\newcommand{\equivClusterThetaAEps}[1]{\equivCluster{\theta}{#1}{a}{\varepsilon_a}}
\newcommand{\representative}[1]{\operatorname{repr}(#1)}
\newcommand{\candidates}[2]{\operatorname{candidates}(#1, #2)}

\newcommandx{\norm}[2][2=2]{\| #1 \|_{#2}}
\newcommand{\indexedSet}[3]{\myVec{#1}_{#2}^{#3}}
\newcommand{\TVD}[1]{ \| #1 \|_{\mathrm{TV}} }
\newcommand{\TVDij}[0]{\TVD{f_i^a-f_j^a}}

\newcommand{\TVDih}[0]{\TVD{f_i^a-f_h^a}}

\newcommand{\partialDerive}[1]{\frac{\partial}{\partial#1}}
\newcommand{\partialDeriveSecond}[1]{\frac{\partial^2}{\partial#1^2}}

\date{}

\title{Iterative Hypothesis Pruning and Distribution-based Early Labeling for Sequential Hypothesis Testing}
\ifBlind
    \author{%
      \IEEEauthorblockN{Anonymous Authors}
      \IEEEauthorblockA{%
        }
    }
\else
    \author{
        \IEEEauthorblockN{George Vershinin, Asaf Cohen, and Omer Gurewitz}
        
        \IEEEauthorblockA{The School of Electrical and Computer Engineering,
                        Ben-Gurion University of the Negev, Israel
                        \newline
                        georgeve@post.bgu.ac.il, \{coasaf, gurewitz\}@bgu.ac.il}
    }
\fi

\begin{document}

\maketitle
\begin{abstract}
\ifPaperAward
    THIS PAPER IS ELIGIBLE FOR THE STUDENT PAPER AWARD.
\fi
We consider the framework of Sequential Hypothesis Testing (SHT), in which a decision maker (DM) selects actions that generate samples from known, action-dependent distributions, while the realized distribution is determined by an unknown true hypothesis. To identify this hypothesis, we adopt the elimination perspective and propose three deterministic, adaptive, multi-iteration algorithms with a common structure, termed $\Phi$, $\Phi$-$\Delta$, and $I$. In each iteration, the DM selects an action and repeatedly applies it to collect samples, after which hypotheses inconsistent with the observed data are eliminated. The algorithms differ in the criterion used to terminate each iteration: $\Phi$ continues until one hypothesis dominates all others; $\Phi$-$\Delta$ first clusters hypotheses whose per-action distributions are close in total variation and then proceeds in the spirit of $\Phi$; $I$ continues until one hypothesis can be safely discarded.

We analyze our algorithms, establishing: (i) controlled error-rates, (ii) controlled sample complexity, (iii) asymptotic optimality, (iv) computational complexity, and (v) NP-hardness of the optimal action-sequence selection for minimal sample complexity.

\end{abstract}
\section{Introduction}
\label{section: Intro}
\HT is a foundational statistical methodology for evaluating competing hypotheses and identifying the one most consistent with observed data.
Its use predates formal statistical theory and spans domains ranging from selecting physical models based on empirical measurements (e.g., estimating the Earth’s circumference or validating quantum-mechanical predictions) to supporting modern scientific practice, particularly in medical research, where it underlies treatment evaluation and clinical diagnosis (e.g., early infectious-disease detection such as COVID-19).

Modern computing systems, including autonomous control mechanisms, anomaly-detection algorithms, quality-control pipelines, and large-scale sensor networks, also rely on \HT to detect events or classify observations.
These applications require decision rules that are reliable, interpretable, and computationally efficient, properties that the \HT framework naturally provides.
The rapid expansion of \ML and \AI has further renewed interest in \HT, as many modern inference and classification tasks remain grounded in, or can be rigorously analyzed through, classical \HT principles.
A \DM typically computes the \LLR (or its non-logarithmic form), compares it against a fixed threshold, and selects the corresponding hypothesis.
For instance, the Neyman-Pearson likelihood-ratio test \cite[Theorem 11.7.1]{CoverThomas2006} chooses the distribution that best explains a given set of samples under prescribed error constraints.

Wald’s seminal work \cite{Wald_1945_SHT} extended the fixed-sample framework to sequential settings, introducing binary \SHT, in which samples are drawn one at a time and the \DM adaptively determines when sufficient evidence has accumulated to make a decision.
The resulting scheme, which we refer to as the \emph{Wald Test}, employs two thresholds, one for each hypothesis, and terminates when the evolving \LLR crosses either threshold.
Wald and Wolfowitz showed that \SHT attains the same error guarantees while requiring a minimal number of samples \cite{Wald_Wolfowitz_1948_SPRT_Optimality}.

The Armitage Test \cite{Armitage1950_SHT_MultipleHypotheses, Bar_Tabrikian2018_Composite_SHT_Single_Source} extends the Wald Test to multi-hypothesis settings by conducting pairwise “tournaments,” in which each newly acquired sample updates concurrent Wald-type comparisons between all hypothesis pairs. The \DM then selects the hypothesis that prevails in these tournaments.
Its asymptotic optimality was established in \cite{Dragalin_etAl_1999_MSPRT_AsympOpt}, and its expected sample complexity was further analyzed in \cite{Dragalin_etAl_2000_MSPRT_MeanSamplesApprox}.

In many practical scenarios, the \DM collects and processes data in real time and can often choose from which data source to draw the next sample based on previous samples.
For example, in medical diagnosis, a physician may select which examination to perform next based on previous examinations, where the distribution of test outcomes for each possible illness depends on the chosen examination.
Similarly, an industrial quality-control system may decide which production line or sensor to inspect next based on prior measurements, and a network administrator may select which routers to probe when detecting and classifying potential cyber-attacks.
Such sequential selection enables more efficient and targeted information acquisition.
In these settings, the \DM must balance two key objectives: achieving high decision accuracy while minimizing the number of samples required to reach a decision, as each additional sample may incur costs such as increased latency.

To incorporate actions, Chernoff \cite{Chernoff1959SequentialHT} extended the Wald Test, giving rise to active \SHT.
In this model, the \DM adaptively selects actions from a predefined set, where each action produces a sample whose distribution depends on the action as well as the underlying hypothesis.
The objective is to choose actions and stopping rules that minimize the expected sample size while satisfying prescribed error probabilities.
This framework established the foundation for modern adaptive \HT and has since been generalized to various multi-hypothesis and controlled-sensing scenarios (e.g., \cite{bessler1960theory, Nitinawarat2013_SHT_Argmax2_KLD_wProofs, Naghshvar_Javidi2013_SHT_DynamicProgramming, Bai_Katewa_Gupta_Huang2015_Stochastic_Source_Selection}).

Several deterministic policies have been explored in recent years.
Some works incorporate \ML or deep learning into sequential testing, either in conjunction with the Wald Test (e.g., \cite{Gurevich2019_EEST, Joseph_DeepLearining1}) or by directly optimizing sample efficiency without invoking it (e.g., \cite{Szostak2024_DeepLearining2, stamatelis2024_DeepLearining3}).
A prominent non-learning action-selection policy is the DGF policy for anomaly detection \cite{Cohen_Zhao2015_SHT_AnomalyDetection, Huang2019_DGF_Heterogeneous, Lambez2022_DGF_wSwitchCost, Citron_Cohen_Zhao2024_DGF_on_Hidden_Markov_Chains} that aims to minimize detection times by probing the processes corresponding to either the largest or second largest accumulated \LLRs.
Although the DGF policy outperforms Chernoff’s approach, its main limitation is that it is tailored to anomaly-detection settings in which actions generate samples from only two distributions, making extensions to general multi-hypothesis models non-trivial.

In this paper, we depart from the previously mentioned conventional approaches by employing an elimination strategy rather than a traditional search method.
Instead of accumulating evidence to identify the correct hypothesis, we use samples to systematically eliminate hypotheses that are “almost surely” incorrect.
The rationale for preferring elimination over the search-for-winner strategy stems from a key limitation in the latter approach: in the search strategy, hypotheses compete against each other simultaneously, with the \DM’s actions and sampling decisions being guided by the underlying hypothesis and its closest hypotheses.
While this approach may be effective with a few hypotheses, it becomes inefficient when handling multiple hypotheses, as separating two closely related hypotheses requires an enormous number of samples.

In contrast, the elimination strategy efficiently discards incorrect hypotheses by focusing on those that are most distinctly different from each other, typically requiring far fewer samples.
The strategy proceeds sequentially, with each new action determined by the remaining hypotheses.
Once only two hypotheses remain, the \DM must determine which of these candidates is correct, with the key advantage that one of them is almost surely the true hypothesis, rather than having to isolate an unknown correct hypothesis from among many possibilities.

Our contributions are the following:
\begin{itemize}
    \item We propose three hypothesis-elimination algorithms sharing a common multi-iteration structure, termed $\Phi$, \PHIDELTA, and $I$ (Iota).
    In each algorithm, the \DM first selects an action based on the \TVDtxt between the sample distributions corresponding to the remaining hypotheses, then repeatedly applies the chosen action to collect samples, after which hypotheses inconsistent with the observed data are discarded.
    This process is repeated until only one hypothesis remains, which is declared as true.
    The algorithms differ in their action-selection rules and iteration-level stopping criteria.
    In $\Phi$, actions are chosen according to the smallest non-zero \TVDtxt among the remaining hypotheses, and an iteration ends when one hypothesis has accumulated \LLR against all others sufficiently large.
    \PHIDELTA speeds up this elimination process by clustering hypotheses by \TVDtxt proximity and allowing several clusters to be discarded simultaneously.
    In this context, we focus on clustering distributions from the same \SEF.
    In $I$, actions are chosen to eliminate some hypothesis as quickly as possible, and an iteration terminates once at least one hypothesis can be confidently ruled out.
    \item We analyze the complexity, sample complexity, error probability, and \ABR of our proposed algorithms.
    Additionally, we prove that the optimal action-sequence selection is NP-hard.
    \item We discuss the effectiveness of the elimination framework, and support this discussion with simulations.
    \item For distributions from the same \SEF, we prove that degenerate clustering yields contiguous parameter clusters, and establish monotonicity properties of tilted likelihoods at the parameter extremes.
\end{itemize}

\section{System Model}
\label{section: system model}

\subsection{Notation}
\label{subsection: notation}
All vectors in this manuscript are underlined (e.g., $\myVec{x}$).
We use $\indexedSet{x}{i}{j}$ as a shorthand notation for $(x_i, x_{i+1}, \dots, x_j)^T$, where $T$ is the transpose operation.
The expectation with respect to some random variable $X$ is denoted as $\expValDist{\cdot}{X}$.
For notational convenience, when computing its first moment, we will omit the subscript in the expectation and write $\expVal{X}$ instead of $\expValDist{X}{X}$.
When computing likelihood ratios or \LLRs, we omit the random variable from the expectation subscript but instead write its underlying \PDF, e.g., write $\expValDist{\log (f(X)/g(X)) }{h}$ for densities $f$, $g$, and $h$.
The \TVDtxt between two densities $f$ and $g$ is denoted as $\TVD{f-g} = \frac{1}{2}\int_{\mathbb{R}} \abs{f(x)-g(x)} dx$.
Their \KLDtxt is denoted as $\KLD{f}{g}$, with the conventions that $0\log\frac{0}{0} = 0$, $\log\frac{a}{0} = 0$ and $a\log\frac{a}{0} = \infty$ for any $a>0$.
Unless explicitly specified (e.g., $\ln$), all logarithms in this manuscript are in base two.
Throughout this paper, we adopt the Bachmann–Landau big-O notation as defined in \cite[Chapter~3]{Cormen2009IntroToAlgo3}.
Whenever the relevant variable is not clear from context, we explicitly indicate the quantity with respect to which the growth is measured.

\subsection{Model}
\label{subsection: model}
We consider a phenomenon that cannot be observed directly, but whose manifestations can be observed.
The phenomenon is characterized by one of a finite set of hypotheses $\mathcal{H} = \{0, 1, \ldots, H-1\}$.
A \DM can perform actions from a finite set $\mathcal{A} = \{1,2,\ldots,\abs{\mathcal{A}}\}$, such as issuing queries or conducting tests.
Upon selecting an action, the \DM observes a sample whose distribution depends on both the underlying hypothesis and the chosen action.
Consequently, different actions may induce different sample distributions for the same hypothesis.
Specifically, each hypothesis-action pair $(h, a)\in\mathcal{H}\times\mathcal{A}$ induces a probability distribution over the sample space denoted by $f_h^a(x) = f(x|h,a)$.
The distributions $\{f_h^a\}_{h, a}$ are assumed to be known by the \DM.
For notational simplicity, we focus on scalar samples, and extension to non-scalar samples is straightforward.

\begin{figure}[!htbp]
    \centering
    \includegraphics[]{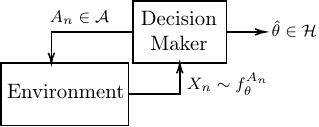}
    \vspace{-8pt}
    \caption{
        System model.
        The \DM is tasked to identify the correct hypothesis indexed by $\theta\in\mathcal{H}$.
        By taking action $A_n$ at time step $n$, the \DM obtains a sample $X_n\sim f_\theta^{A_n}$.
    }
    \label{fig: model}
\end{figure}

We assume no prior knowledge with respect to the underlying hypothesis.
Namely, the prior probability of hypothesis $i$ (denoted by $\hII$) is $\prob{\theta = i} = \frac{1}{H}, \quad \forall i \in \mathcal{H}$.
We further assume that the samples are independent and identically distributed conditioned on the action taken, with the distribution parameters depending on the underlying true hypothesis. 
The decision made is given by $\hat{\theta}\in\mathcal{H}$, i.e., $\hat{\theta} = i$ implies that $\hII$ is declared as true.
Figure \ref{fig: model} visualizes the model.

We make additional assumptions:
\begin{enumerate}[label=(A\arabic*)]
    \item (Separation) For any action $a\in \mathcal{A}$, there are at least two hypotheses $i$, $j\in\mathcal{H}$ with $\TVDij > 0$.
    \label{assumption: separation}
    \item (Validity) For all $i, j\in\mathcal{H}$, there is at least one action $a\in\mathcal{A}$ with $\TVDij > 0$.
    \label{assumption: validity}
    \item (Finite \LLR Variance) There exists $0<\Xi<\infty$ such that $\expValDist{\left(\log\frac{f_i^a(X)}{f_j^a(X)}\right)^2}{f_{i}^a} < \Xi$ for any $i, j\in\mathcal{H}$ and $a\in\mathcal{A}$.
    \label{assumption: finite LLR variance}
\end{enumerate}
The Separation assumption ensures that each action separates at least two hypotheses.
The Validity assumption ensures that all hypotheses are distinguishable.
The Finite \LLR Variance assumption, first introduced by Chernoff \cite{Chernoff1959SequentialHT}, implies that the \KLDstxt are finite, which in turn implies that under each action, the support is shared.
Note that the sample support may not be shared among different actions.

Although the first two assumptions are not necessary, they ensure that the model is both simple and interesting.
Specifically, if the Separation assumption does not hold, some actions cannot distinguish between any hypotheses, and applying them is not only uninformative but also meaningless.
If the Validity assumption does not hold, then the indistinguishable hypotheses (under all available actions) can be treated as one until the \DM retires.
In this case, the \DM must draw one of them randomly and declare it as true, bounding the error probability away from zero.

Let $\Psi$ be the source selection process generating the action sequence $\{A_n\}_{n=1}^\infty$.
Like many other works in the literature, e.g., \cite{Chernoff1959SequentialHT, Nitinawarat2013_SHT_Argmax2_KLD_wProofs, Cohen_Zhao2015_SHT_AnomalyDetection, Gafni2023_CompositeHT, Citron_Cohen_Zhao2024_DGF_on_Hidden_Markov_Chains}, we focus on the Bayesian approach.
Namely, let $\Gamma\triangleq(\Psi, \hat{\theta})$ be an admissible strategy for the \SHT aiming at minimizing the \ABR:
\begin{align}
    \label{eq: Bayes Risk}
        \min_{\Gamma} \{\delta\expVal{N | \Gamma} + p_e(\Gamma)\}
    ,
\end{align}
where $N|\Gamma\in\mathbb{N}$ is the stopping time of $\Gamma$ (before delivering the decision given by $\hat{\theta}$), $\delta\in(0, 1)$ is a regularizer interpreted as the sample cost, and $p_e(\Gamma)\triangleq\expValDist{\mathbb{P}(\hat{\theta} \neq \theta) \middle| \Gamma}{\theta}\in(0, 1)$ is the average error probability of policy $\Gamma$.

Eq. \eqref{eq: Bayes Risk} weights two key components of the problem: the error probability, $p_e$, and the product of the expected number of samples $\expVal{N|\Gamma}$ times $\delta$, which corresponds to the expected total cost incurred until termination.

Notably, the \ABR does not exceed one only if $\delta\expVal{N|\Gamma}\in(0, 1)$, motivating the interest in algorithms whose expected number of samples is sublinear in $1/\delta$, yet enjoy small error probability, e.g., $p_e\leq \delta$ when $\delta$ is sufficiently small.
Particularly, when $\expVal{N|\Gamma} = \littleO{1/\delta}$ and $p_e = \bigO{\delta}$ where the little-O and big-O are with respect to $\delta\to 0$, the \ABR vanishes when $\delta\to0$.

\section{Preliminaries}
\label{section: Preliminaries}
This section reviews two canonical \SHT procedures used throughout the paper and provides a brief overview of the \SEF, which serves as the setting for part of our analysis.

\subsection{The Wald Test}
\label{subsection: Wald}
The Wald Test, also known as the \SPRT, is a canonical sequential test designed for two-hypothesis setting ($\mathcal{H} = \{0,1\}$) with a single action ($|\mathcal{A}| = 1$). 

The Wald Test relies on the fact that, as the number of samples grows, the \LLR between the two candidate hypotheses increasingly favors the true hypothesis.
Specifically, at each timestep, the \DM invokes the action, collects an additional outcome sample, and updates the accumulated \LLR based on all previously observed samples.
The \emph{pairwise accumulated \LLR} between $H_0$ and $H_1$ after collecting $n$ samples is
\begin{align*}
    L_{01}(\indexedSet{x}{1}{n})
    \triangleq
    \sum_{t=1}^n \log \frac{f_0(x_t)}{f_1(x_t)} .
\end{align*}

The \DM compares the accumulated \LLR to two decision thresholds.
If the \LLR exceeds the upper threshold, the \DM stops and declares $H_0$ to be true; if it falls below the lower threshold, the \DM stops and declares $H_1$ to be true.
Otherwise, the \DM continues sampling.
In this paper, we assume no a priori knowledge and no bias toward either hypothesis, and therefore adopt the symmetric Wald Test, in which the thresholds are set to $\pm \gamma$ for some $\gamma>0$.
Under these conditions, symmetric thresholds yield a balanced stopping rule that treats false positives and false negatives equivalently.
Moreover, in the multi-hypothesis setting considered here, symmetric thresholds enable a uniform thresholding scheme across all pairwise comparisons.


\subsection{The Armitage Test}
\label{subsection: Armitage}
The Armitage Test, widely known as the \MSPRT, extends the Wald Test to settings where the number of hypotheses exceeds two.

At each time step, similar to the Wald Test, the \DM collects an additional observation.
Based on the collected samples, the \DM updates the accumulated \LLRs for each pair of hypotheses and checks whether any hypothesis is sufficiently dominant over the others.
In particular, for a given threshold $\gamma$, \textit{$\hII$ }is said to \emph{lose} against $\hJJ$ in the pairwise comparison if $L_{ij}(\indexedSet{x}{1}{n}) \leq -\gamma$, and to \emph{win} against $\hJJ$ if $L_{ij}(\indexedSet{x}{1}{n}) \geq \gamma$.

The Armitage Test continues until there exists an index $i$ such that the accumulated \LLRs of $\hII$ against every competing hypothesis exceed $\gamma$, at which point $\hII$ is declared as true.
Otherwise, the \DM continues sampling.

\subsection{The Scalar Exponential Family}
\label{subsection: SEF}
Several of our analytical results concern the rich class of distributions known as the \SEF, which includes many fundamental distributions, such as the Bernoulli, geometric, binomial (with a known number of trials), normal (with known variance), gamma, Poisson, and exponential distributions.

The \SEF consists of all probability density (or mass) functions that can be expressed in the form
\begin{align}
    \label{eq: SEF form}
    f(x; \eta) = \zeta(x)\exp\left\{ \eta T(x) - A(\eta) \right\}
    ,
\end{align}
where $\zeta(x)$, $T(x)$, and $A(\cdot)$ are known functions and $\eta$ is the distribution parameter.
In particular, $\zeta(x)$ is a non-negative baseline function that is independent of the parameter, $T(x)$ is the sufficient statistic of the distribution, and $A(\cdot)$ is the log-partition function that ensures that $f$ integrates (or sums) to one.
Additionally, $A(\cdot)$ is convex (by Holder’s Inequality), and, throughout this paper, is assumed to be twice differentiable.


\section{Sequential Hypothesis Elimination Framework}
\label{section: Policy}
In this section, we introduce the proposed sequential hypothesis elimination framework.
In contrast to classical approaches that continue sampling until a single hypothesis dominates all others, the proposed methods progressively discard hypotheses until only one remains.
The action-selection policy is adaptive and aims to minimize the expected number of samples required to eliminate hypotheses that are almost surely incorrect, thereby ensuring identification of the true hypothesis with a prescribed reliability level.

\subsection{The \texorpdfstring{$\Phi$}{PHI} Algorithm}
\label{subsection: generic multi-stage}
The Armitage Test (Section \ref{subsection: Armitage}, \cite{Armitage1950_SHT_MultipleHypotheses}) is simple, computationally efficient, and straightforward to implement.
However, it is inherently designed for a single-action setting, in which a fixed action induces distinct distributions across competing hypotheses.
Additionally, in many practical scenarios, the \DM can choose among multiple actions, each inducing a different observation model under the competing hypotheses, and the \DM must also determine which action to select.

A na\"{\i}ve extension of the Armitage Test to the multi-action setting, in which the \DM selects a single action and applies a standard sequential test, may perform poorly when there is no single action that induces strong statistical separation between all hypotheses.
All the more so, when under a given action, there are multiple hypotheses that induce identical distributions, rendering them indistinguishable regardless of the number of collected samples.
In such cases, the Armitage Test cannot discriminate between these hypotheses unless a different action is taken.
The \PHI, or $\Phi$, algorithm, which we describe next, extends the Armitage Test to the multi-action setting.
It is an iterative algorithm that addresses both challenges: (i) selecting an appropriate action at each iteration, and (ii) ensuring reliable discrimination between hypotheses that induce identical distributions under certain actions.

The key design principle of $\Phi$ is to select, at each iteration, an action that maximizes the worst-case statistical separation between the remaining candidate hypotheses.
To formalize this notion, we require a measure of separation between hypotheses.
Importantly, this measure must be \emph{symmetric}, so that the separation between $\hII$ and $\hJJ$ does not depend on which hypothesis is taken as the reference.
Symmetry ensures that elimination decisions are driven by intrinsic statistical distinguishability rather than by an arbitrary ordering of hypotheses.
We therefore adopt the \TVDtxt to quantify the separation between the distributions induced by different hypotheses under a given action.
Specifically, at each iteration, given the current hypothesis set $\mathcal{U}\subseteq\mathcal{H}$, the $\Phi$ algorithm selects an action according to
\begin{align*}
    a^* = \argmax{
    \left\{
        \min_{\substack{i,j\in\mathcal{U}\\ \TVDij > 0}} \TVDij
    \right\}
    }{a \in \mathcal{A}}
    ,
\end{align*}
i.e., the action that maximizes the worst-case (non-zero) pairwise \TVDtxt separation among the remaining hypotheses.

To address the second challenge, namely, indistinguishable hypotheses under certain actions, we group hypotheses that are indistinguishable under the selected action into equivalence classes and treat each such class as a single aggregate alternative.
Formally, define the equivalence class of a hypothesis $H_i$ under action $a$ as the set of all hypotheses in $\mathcal{U}$ that induce the same distribution under $a$:
\begin{align*}
    \equivClassIA{\mathcal{U}}
    \triangleq
    \left\{
        h \in \mathcal{U}
        \; : \;
        \TVDih = 0
    \right\}
    .
\end{align*}
For each equivalence class under action $a$, we define a single representative to characterize the class under that action. Without loss of generality, we choose the representative of each equivalence class to be the hypothesis with the smallest index, i.e.,
\begin{align}
    \nonumber
    \representative{ \equivClassIA{\mathcal{U}} }
    &\triangleq
    \min\{\equivClassIA{\mathcal{U}}\}
\end{align}
Let $\candidates{\mathcal{U}}{a}$ denote the set of all representatives from $\mathcal{U}$ under action $a$. Formally,
\begin{align}
    \nonumber
    \candidates{\mathcal{U}}{a}
    \triangleq
    \bigcup_{h\in\mathcal{U}}
    \representative{ \equivClass{h}{a}{\mathcal{U}} }
    .
\end{align}
An equivalence class may contain a single hypothesis if no other hypothesis induces the same distribution under action $a$.
Assumptions \ref{assumption: separation} and \ref{assumption: validity} ensure that there is no action $a$ for which $\abs{\candidates{\mathcal{H}}{a}} = 1$.
Moreover, if all output distributions induced by action $a$ are distinct, then $\abs{\candidates{\mathcal{H}}{a}} = \abs{\mathcal{H}} = H$.
Accordingly, for each action $a$, the number of equivalence classes lies between $2$ and $H$.



After selecting an action, the $\Phi$ algorithm applies the Armitage Test to the equivalence class representatives.
Once a single representative wins the competition, the $\Phi$ algorithm discards all hypotheses belonging to the remaining equivalence classes and repeats the procedure: it recomputes the equivalence classes of the surviving hypotheses under the available actions, selects the action with the largest inter-class \TVDtxt, and applies the Armitage Test again.
This process continues until only a single hypothesis remains.
Note that the procedure may terminate after a single iteration if the winning equivalence class in the first competition corresponds to a singleton, in which case the $\Phi$ directly identifies the correct hypothesis.

A pseudocode description of the $\Phi$ algorithm is provided in Algorithm~\ref{alg: Multi-Stage SHT}. In Line~\ref{alg line: init hAlive}, the set of \emph{alive} hypotheses, denoted by $\hAlive$, is initialized to include all hypotheses. The algorithm then proceeds iteratively, reducing the number of candidate hypotheses at each iteration until a single hypothesis remains, which is declared the true one.

Lines~\ref{alg line: round start}–\ref{alg line: round end} (the outer \texttt{while} loop) correspond to these iterations. At the beginning of each iteration, the action $a$ that maximizes the minimum distance, i.e., provides the best separation, between non-equivalent alive hypotheses is selected (Line~\ref{alg line: compute action}). Line~\ref{alg line: compute dist hypotheses} determines the set of contestants for the current iteration from among the alive hypotheses.

The selected action $a$ is then repeatedly applied to collect samples and update the accumulated \LLRs until a hypothesis $i^*$ emerges as a winner against all others (Lines~\ref{alg line: LLR update start}–\ref{alg line: LLR update stop}).
This winner $i^*$ is defined as the hypothesis whose equivalence class achieves a minimal accumulated \LLR exceeding the threshold $\gamma$ against every other equivalence class.

Finally, $\hAlive$ is updated to the equivalence class of $i^*$ under the selected action $a$ (Lines \ref{alg line: get equivalent of winner}–\ref{alg line: update alive hypotheses}), and the procedure repeats.

\begin{algorithm}[!htbp]
    \caption{ The $\Phi$ Algorithm }
    \label{alg: Multi-Stage SHT}
    \begin{algorithmic}[1]
        \State $\hAlive\gets \mathcal{H}$          \Comment{Initialize alive hypotheses} \label{alg line: init hAlive}
        \State $L_{ij}\gets 0\ \forall i \neq j\in \mathcal{H}$           \Comment{Initialize \acp{LLR}}  \label{alg line: init LLRs}
        \While{ $\abs{\hAlive} \geq 2$ }
            \State $a^*\gets \displaystyle\argmax{ \min_{ \substack{i,j\in\hAlive \\ j\not\in\equivClassIA{\hAlive}} } \TVDij }{a\in \mathcal{A}}$ \label{alg line: compute action}
            \label{alg line: round start}
            \State $\tilde{H} \gets \candidates{\hAlive}{a^*}$   \label{alg line: compute dist hypotheses}            
            \While{ $\not\exists i\in \tilde{H}$: $L_{ij} \geq \gamma\ \forall j\neq i\in\tilde{H}$ }
                \label{alg line: LLR update start}
                \State acquire a single sample, $x$, by applying action $a^*$ \label{alg line: acquire data}
                \State $L_{ij} \gets L_{ij}
                +
                \log\frac{ f_i^{a^*}(x) }{ f_j^{a^*}(x) }$ for all $j\neq i\in \Tilde{H}$
            \EndWhile
            \label{alg line: LLR update stop}
            \State $\hat{H} \gets \equivClass{i^*}{a^*}{\hAlive}$  \Comment{$i^*$ has $L_{i^*j}\geq \gamma\ \forall j$} \label{alg line: get equivalent of winner}
            \State $\hAlive \gets \hAlive \cap \hat{H}$    \Comment{Update alive hypotheses}  \label{alg line: update alive hypotheses}
            \label{alg line: round end}
        \EndWhile
        \State return $\hAlive$
    \end{algorithmic}
\end{algorithm}

\subsection{Per-Action Hypothesis Clustering: The \texorpdfstring{\PHIDELTA}{PHIDELTA} Algorithm}
\label{subsection: clustering step}
In many practical settings, distinct hypotheses can induce output distributions that are nearly indistinguishable under a given action (e.g., due to different noise realizations), as reflected by small \TVDtxt and \KLDtxt.
Since, for a fixed action, the expected number of samples required to distinguish the true hypothesis from its closest competitor scales inversely with \KLDtxt (see, e.g., \cite[Theorem~4.1]{Dragalin_etAl_1999_MSPRT_AsympOpt}), small \KLDtxt values can lead to prohibitively large sample complexity.
Consequently, the $\Phi$-algorithm, which groups, for each action, only hypotheses that induce \emph{identical} distributions into the same equivalence class, may also incur prohibitively large sample complexity.

The \PHIDELTA algorithm addresses this limitation by allowing per-action equivalence classes to include hypotheses whose induced distributions lie within a prescribed neighborhood, rather than requiring exact equality.
This relaxation increases the effective separation between distinct classes, thereby reducing the number of samples needed to reliably eliminate competing classes.

The \PHIDELTA algorithm follows a structure similar to that of the $\Phi$-algorithm.
At each phase, the \DM identifies, for each action, the equivalence classes and their representatives, selects an action, performs an Armitage test among the class representatives, and eliminates all hypotheses belonging to the losing classes.
The key distinction is that, in \PHIDELTA, equivalence classes are defined through a statistical proximity criterion rather than exact equality of the induced distributions.
Consequently, the selection of suitable representatives requires additional care.
We first describe the clustering mechanism and then discuss the selection of cluster representatives.

While the discussion highlights general principles underlying clustering and representative selection, our primary focus is on the specific clustering mechanism adopted for \PHIDELTA in this work.
We do not claim that this mechanism is optimal; rather, it serves as a concrete instantiation of the general approach.

\subsubsection{The Clustering Mechanism}
As in the $\Phi$-algorithm, \PHIDELTA equivalence classes are defined per action due to the dependence of the sampling distribution on the selected action.
In the sequel, we focus on the clustering mechanism associated with a fixed action.

As discussed earlier, the sample complexity is governed by the minimum \KLDtxt between competing hypotheses, suggesting that the \KLDtxt is a natural choice for clustering.
However, the \KLDtxt is not a metric, as it is neither symmetric nor does it satisfy the triangle inequality.
In particular, its asymmetry may lead to pathological situations in which $\hII$ is deemed close to $\hKK$, while $\hKK$ is not deemed close to $\hII$.
Such behavior is undesirable, as clustering requires a mutual notion of proximity.

Motivated by Pinsker's inequality, which ensures that small \KLDtxt implies small \TVDtxt \cite[Lemma~11.6.1]{CoverThomas2006}, we adopt the \TVDtxt as the clustering metric. Unlike the \KLDtxt, the \TVDtxt is a metric and therefore provides a symmetric, well-defined notion of proximity, yielding consistent equivalence classes.

Various clustering strategies may be employed, differing in computational complexity and in their effectiveness at facilitating hypothesis elimination.
In the sequel, we elaborate on the clustering algorithm adopted by \PHIDELTA.

Adopting the \TVDtxt as the distance metric enables the use of density-based clustering methods in metric spaces (e.g., \cite[Chapter~5]{Aggarwal_Reddy_2013_ClusteringBook}), which group hypotheses according to proximity, require no prior knowledge of the number of clusters, and are flexible with respect to the underlying distance.
In particular, \PHIDELTA employs a degenerate variant of the Density-Based Spatial Clustering of Applications with Noise (DBSCAN) algorithm \cite{Ester_etAl_1996_DBSCAN, Aggarwal_Reddy_2013_ClusteringBook}.

Specifically, for each action $a$, the algorithm defines a neighborhood radius $\varepsilon_a$ and groups hypotheses into clusters according to $\varepsilon_a$-connectivity.
Two hypotheses are said to be $\varepsilon_a$-connected if there exists a finite sequence of hypotheses linking them such that the \TVDtxt between every consecutive pair in the sequence is at most $\varepsilon_a$.
A cluster is then defined as a maximal set of pairwise $\varepsilon_a$-connected hypotheses.
Note that this construction yields a unique partition into clusters.

Consequently, for every non-singleton cluster, each hypothesis has at least one other hypothesis in the same cluster whose \TVDtxt is at most $\varepsilon_a$, and hypotheses that belong to different clusters are not $\varepsilon_a$-connected; in particular, their pairwise \TVDtxt exceeds $\varepsilon_a$.

The hypotheses clusters per action can be formalized as follows: 
For each action $a \in \mathcal{A}$ and a proximity threshold $\varepsilon_a > 0$, define an undirected graph $G_a = (\mathcal{H}, E_a)$ whose vertex set is the hypothesis set $\mathcal{H}$, and where an edge $(i, j) \in E_a$ exists if and only if $\TVDij \leq \varepsilon_a$.
We say that two hypotheses $i, j \in \mathcal{H}$ are $\varepsilon_a$-connected if they belong to the same connected component of $G_a$. The \emph{\PHIDELTA clusters} under action $a$ are defined as the connected components of $G_a$. That is, each cluster $\mathcal{C} \subseteq \mathcal{H}$ is a maximal set of hypotheses such that for every pair $i, j \in \mathcal{C}$, there exists a finite sequence $\{l_k\}_{k=1}^m \subseteq \mathcal{C}$ with $l_1 = i$ and $l_m = j$ satisfying
\begin{align}
    \TVD{f_{l_k}^a - f_{l_{k+1}}^a} \leq \varepsilon_a, \quad \forall k = 1, \dots, m-1.
\end{align}
Equivalently, the collection of clusters forms a partition of $\mathcal{H}$ into disjoint equivalence classes under the relation of $\varepsilon_a$-connectivity.
We hereafter denote by $\equivClusterIAEps{\mathcal{U}}$ the cluster of $\hII$ that consists of hypotheses from a given hypothesis set $\mathcal{U}\subseteq\mathcal{H}$.

The clustering procedure can be described as follows.
For each action, initialize a cluster with an arbitrary hypothesis and include in it all hypotheses whose \TVDtxt from it is at most $\varepsilon_a$.
Then, iteratively examine the remaining hypotheses and add any hypothesis that lies within \TVDtxt at most $\varepsilon_a$ from at least one hypothesis already in the cluster.
Each time a new hypothesis is added, all hypotheses not yet assigned to the cluster are re-examined, including those previously excluded.
This process continues until no further hypotheses can be added.
Once the cluster is complete, select a hypothesis that has not yet been assigned to any cluster and repeat the procedure.
The process terminates when all hypotheses have been assigned to clusters.

The proximity threshold $\varepsilon_a$ plays a central role in determining the clustering structure under each action, as it governs both the cluster cardinalities and the separation between distinct clusters.
In particular, it induces a fundamental trade-off between the sample complexity per iteration and the discrimination granularity, which in turn determines the number of required iterations.

When $\varepsilon_a$ is large, the hypothesis space collapses into a small number of clusters, possibly even a single cluster, with relatively large separation between them.
This leads to a small sample complexity per iteration (in the extreme case of a single cluster, no discrimination is required at that iteration), but typically necessitates multiple iterations to fully resolve the true hypothesis.
Conversely, as $\varepsilon_a$ vanishes, the clustering becomes increasingly fine-grained, yielding many clusters with small inter-cluster separation.
In this regime, the per-iteration sample complexity increases, while the number of iterations decreases; in the limit, the procedure reduces to the original $\Phi$ algorithm.

Ideally, the threshold $\varepsilon_a$ should be selected dynamically for each action to adapt to the current set of surviving hypotheses.
For instance, one may begin with a relatively large $\varepsilon_a$ and gradually decrease it as the hypothesis set shrinks, ensuring that at least two distinct clusters remain separable.
In particular, when only two hypotheses remain, and their \TVDtxt is small, $\varepsilon_a$ should be chosen to ensure their separation under that action. 
In this paper, for simplicity, we adopt a fixed $\varepsilon_a$ for all actions and reduce it only when necessary, namely, when it fails to separate hypotheses that do not induce identical output distributions.

\subsubsection{The Cluster Representatives}
The purpose of clustering is to enable the elimination of an entire set of hypotheses based solely on comparisons between its representative(s) and those of competing clusters.
Thus, the representative(s) of a cluster must serve as reliable proxies for all hypotheses it contains.
A cluster may be represented either by a single representative or by multiple representatives, which may be actual hypotheses within the cluster or suitably constructed virtual ones.

When a single representative is used, it must be selected so that, if it is eliminated in competition with representatives of other clusters, the entire cluster can be safely discarded.
When multiple representatives are employed, the elimination rule requires additional design choices, e.g., determining whether eliminating a single representative is sufficient to discard the cluster, whether all representatives must be eliminated, or whether an intermediate criterion should be used.
In general, the design of the representatives and the associated elimination rules should balance two objectives: faithfully capturing the hypotheses within the cluster while maximizing the efficiency of subsequent iterations.

In the sequel, we focus on the \SEF setting.
We assume that different actions may induce different \SEF families (e.g., one action may induce an exponential distribution, whereas another may induce a Poisson distribution).
However, for any fixed action, all hypotheses induce distributions from the same \SEF family, differing only in their parameter values.

Within the \SEF framework, each hypothesis is characterized by a single parameter.
Thus, the hypotheses can be represented as points on the real line according to their parameter values.
Under the proposed clustering mechanism, the resulting clusters are contiguous in this parameter space.
Specifically, if two hypotheses with parameter values $\eta_i$ and $\eta_j$ belong to the same cluster, then every hypothesis whose parameter lies in the interval between $\eta_i$ and $\eta_j$ also belongs to the same cluster (see Lemma \ref{lemma: exp family clustering properties}).
Accordingly, each cluster can be represented by its two boundary hypotheses, namely, those with the smallest and largest parameter values within the cluster.

\begin{figure}[!htbp]
    \centering
    \includegraphics[scale=0.3]{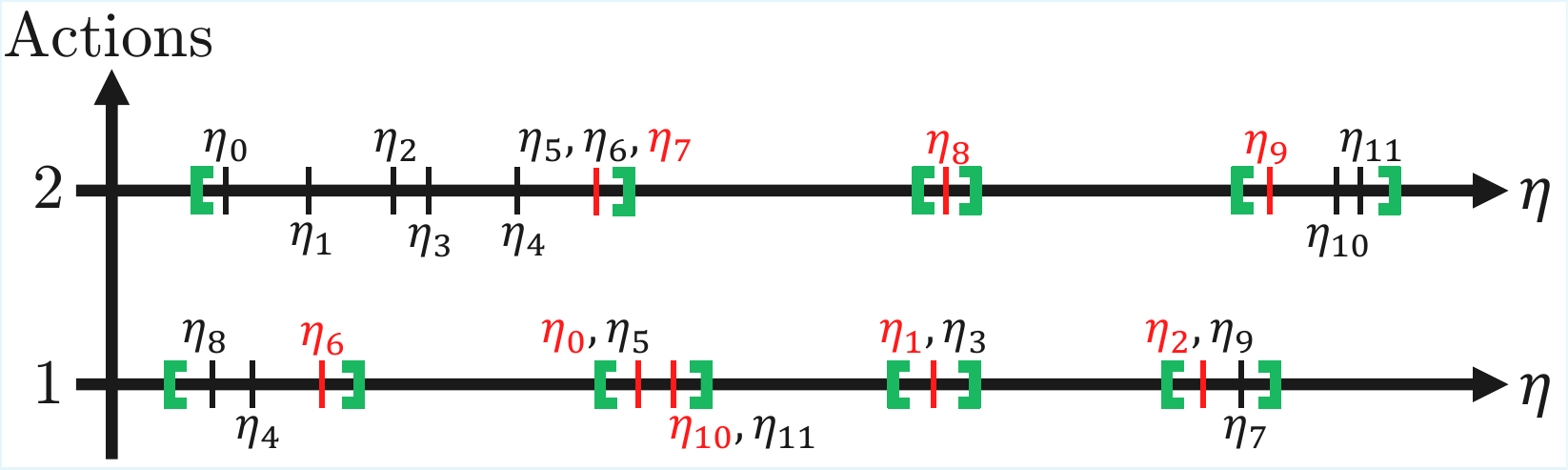}
    \vspace{-8pt}
    \caption{
        The \TVDtxt-induced clusters in the parameter space $\eta\in\mathbb{R}$ for two actions when the sample distribution comes from the \SEF.
        The hypotheses are sorted by their distribution parameter ($\eta_i$).
        Markers on the x-axis in red highlight the distributions closest to the cluster boundaries (green) that will compete in Wald Tests.
        For example, if action 1 is taken and hypotheses $H_0$ and $H_3$ are to compete, then their corresponding representatives in their respective Wald Test are $H_{10}$ and $H_1$, respectively. For its competition with $H_8$, which is represented by $H_6$ under action 1, $H_0$ represents itself. 
    }
    \label{fig: equivalent hypotheses}
\end{figure}

Figure \ref{fig: equivalent hypotheses} illustrates this property for two actions.
The square brackets indicate the cluster boundaries.
Under action 1, the hypothesis set $\mathcal{H}=\intSet{0}{11}$ is partitioned into the clusters $\{4,6,8\}$, $\{0,5,10,11\}$, $\{1,3\}$, and $\{2,7,9\}$, with representatives $\{6,8\}$, $\{0,10\}$, $\{1,3\}$, and $\{2,7\}$, respectively.
Under action 2, $\mathcal{H}$ is partitioned into the clusters $\intSet{0}{7}$, $\{8\}$, and $\{9, 10, 11\}$, with representatives $\{0,7\}$, $\{8\}$, and $\{9,11\}$, respectively.
The representatives are highlighted in red.
Furthermore, when clusters compete, it suffices to compare only the pair of boundary representatives that are closest in parameter value (again, see Lemma \ref{lemma: exp family clustering properties}).
As a result, each cluster needs to compete only with its two adjacent clusters; those immediately to its left and right on the parameter axis.

These comparisons induce a monotone elimination rule.
If a cluster representative defeats that of an adjacent cluster, then not only is the losing cluster eliminated, but so are all clusters lying beyond it in the same direction along the parameter axis.
Specifically, if the smallest-parameter representative of a cluster defeats the representative of the adjacent cluster with smaller parameter values, then all clusters with smaller parameter values can be safely discarded.
By symmetry, if the largest-parameter representative defeats that of the adjacent cluster with larger parameter values, then all clusters with larger parameter values can likewise be eliminated.

For example, in Figure \ref{fig: equivalent hypotheses}, consider action 1.
If $\eta_1$ defeats $\eta_{10}$, then the clusters $\{4,6,8\}$ and $\{0,5,10,11\}$, which lie to the left of the winning cluster $\{1, 3\}$, can both be discarded.
Conversely, if $\eta_{10}$ defeats $\eta_1$, then the clusters $\{1,3\}$ and $\{2,7,9\}$, which lie to the right of the winning cluster $\{0,5,10,11\}$, can be discarded.

\subsection{The \texorpdfstring{$I$}{I} Algorithm}
\label{subsection: IOTA}
We move to introduce the last elimination algorithm, the \IOTA algorithm, hereafter referred to as the $I$ algorithm.
The $I$ algorithm follows the same spirit as Algorithm \ref{alg: Multi-Stage SHT}.
Specifically, at each iteration, an action is selected, and the Wald Test is run between the alive hypotheses until at least one hypothesis can be removed from the candidate pool.
This process repeats until a single hypothesis remains.

In the $I$ algorithm, the actions are selected to enable the fastest elimination of hypotheses that are almost surely incorrect.
Accordingly, the selected action should induce the strongest separation among the remaining candidate hypotheses, i.e., at each iteration, given the current hypothesis set $\mathcal{U}\subseteq\mathcal{H}$, the $I$ algorithm selects an action according to
\begin{align*}
    a^* = \argmax{ \left\{\max_{i,j\in\mathcal{U}} \TVDij \right\} }{a \in \mathcal{A}}
\end{align*}
that is, the action that maximizes the largest best-case \TVDtxt separation between a single hypothesis and the remaining hypotheses.
Using the selected action $a^*$, the algorithm applies pairwise Wald Tests among the hypotheses in $\mathcal{U}$.

Once a hypothesis loses to any other hypothesis, it is immediately pruned from $\mathcal{H}$ and no further comparisons involving that hypothesis are performed. The procedure is then repeated in a new iteration with the reduced hypothesis set until a single hypothesis remains.


\section{Performance Analysis}
\label{section: Performance Analysis}
In this section, we analyze the performance of the proposed algorithms. 
For each algorithm, the analysis consists of:
(i) bounding the error probability;
(ii) computing the sample complexity;
(iii) establishing asymptotic optimality as $\delta\to 0$;
(iv) analyzing the space and average runtime complexity.
Additionally, we show that computing the action sequences that minimize sample complexity is NP-hard.

\subsection{Bounding the Error Probability}
\label{subsection: PE Analysis}
We begin by showing how the decision threshold $\gamma$ controls the overall error probability of the $\Phi$ algorithm:
\begin{lemma}
    \label{lemma: Phi pe bound}
    $p_e(\Phi)\leq (H-1)\times2^{-\gamma}$.
\end{lemma}
\begin{IEEEproof}
    The proof relies on the fact that the $\Phi$ algorithm relies on successive Armitage Tests between equivalence classes’ representatives under each iteration’s selected action, where the number of contending equivalence classes per iteration is varied based on the hypotheses already being discarded on previous iterations and the new selected action.
    Utilizing the union bound both for bounding the error probability per iteration and for bounding the error probability for multiple iterations yields the bound.
    \ifShowSupp
        The detailed proof is given in Appendix \ref{lemma: Phi pe bound proof}. 
    \fi    
\end{IEEEproof}

We next consider \PHIDELTA.
In contrast to the $\Phi$ algorithm, where at each iteration the representative of an equivalence class induces the same distribution as all hypotheses in the class under the selected action, in \PHIDELTA, a cluster representative need not share the same distribution as its cluster members.
In particular, the distribution induced by the representative of the cluster containing the true hypothesis $\theta$ may differ from that induced by $\theta$ under the selected action.
Consequently, elimination decisions based on the \LLR of the cluster representative may lead to the erroneous removal of the entire cluster, including the true hypothesis.

Although \PHIDELTA operates with a finite number of samples per phase, we begin by analyzing the asymptotic behavior of the accumulated \LLR within a single phase under a fixed action $a$, as this provides fundamental insight into its error behavior.
Specifically, consider the accumulated \LLR between two arbitrary hypotheses, $\hII$ and $\hJJ$, neither of which is necessarily equal to $H_\theta$.
Since samples are generated according to $\theta$, the normalized accumulated \LLR converges almost surely \cite[Theorem~16.8.1]{CoverThomas2006} to
\begin{align}
    \nonumber
    \frac{1}{n} L_{ij}
    &\xrightarrow{n \to \infty}
    \expValDist{ \log\frac{ f_i^a(X) }{ f_j^a(X) } }{ f_\theta^a}
    \\
    \nonumber
    &=
    \expValDist{ \log\left(\frac{ f_i^a(X) }{ f_j^a(X) }\times\frac{f_\theta^a(X)}{f_\theta^a(X)}\right) }{ f_\theta^a}
    \\
    \nonumber
    &=
    \expValDist{ \log\frac{ f_\theta^a(X) }{ f_j^a(X) } }{ f_\theta^a}
    -
    \expValDist{ \log\frac{ f_\theta^a(X) }{ f_i^a(X) } }{ f_\theta^a}
    \\
    \nonumber
    &=
    \KLDthetaj - \KLDthetai
    \\
    \label{eq: KLD Difference}
    &
    \triangleq \DKLD{\theta}{j}{i}{a},
    \qquad \forall i, j \in \mathcal{H}
    .
\end{align}
Accordingly, the sign of $\DKLD{\theta}{j}{i}{a}$ determines which of the two hypotheses is closer to the true distribution in the \KLDtxt sense.
In particular, if $\DKLD{\theta}{j}{i}{a} > 0$, i.e., $\KLDthetaj > \KLDthetai$, then $\hJJ$ is less compatible with the samples generated by $H_\theta$ than $\hII$, and vice versa $\DKLD{\theta}{j}{i}{a} < 0$ implies that $\hJJ$ is more compatible with the samples generated by $\theta$ than $\hII$.

This convergence holds for every pair of hypotheses, and in particular for the cluster representatives.
Accordingly, consider any pairwise competition between cluster representatives in which one of the clusters contains the true hypothesis.
If the representative of this cluster is closer, in the \KLDtxt sense, to the true hypothesis than the competing representative, then, given sufficiently many samples, the sign of the accumulated \LLR will correctly favor it, leading to the elimination of the competing cluster, which therefore cannot contain the true hypothesis.
Consequently, if the cluster formation and the choice of representatives preserve the \KLDtxt ordering, so that the representative of the cluster containing the true hypothesis is closer, in the \KLDtxt sense, to the true hypothesis than any other cluster representative, then the cluster whose representative prevails over all others must contain the true hypothesis.

Since \PHIDELTA operates with a finite number of samples per iteration, and terminates an iteration once a single representative defeats all others, i.e., $\hII$ is declared the winner when $L_{ij} \geq \gamma\; \forall j \neq i$, the convergence of the accumulated \LLR to the true \KLDtxt difference with respect to the true hypothesis cannot be guaranteed.
Consequently, errors may occur, and in particular, the cluster containing $\theta$ may be incorrectly discarded. The following proposition bounds the probability of this event under the assumption that the representative of the cluster containing $\theta$ is closer, in the \KLDtxt sense, to $\theta$ than any other cluster representative under the selected action.

\begin{proposition}
    \label{prop: PHIDELTA pe bound Full}
    Consider a single phase of the \PHIDELTA algorithm under a fixed action $a$.
    Suppose that the representative $k(\theta)$ of the cluster containing the true hypothesis $\theta$ satisfies $\KLD{f_\theta^a}{f_{k(\theta)}^a} < \KLD{f_\theta^a}{f_i^a},\ \forall i \notin \equivCluster{\theta}{\mathcal{U}}{a}{\varepsilon_a}$.
    Then, for any $i \notin \equivCluster{\theta}{\mathcal{U}}{a}{\varepsilon_a}$, the probability that $k(\theta)$ loses to $i$ is upper bounded by $\mathbb{P}_\theta\big(i \text{ defeats } k(\theta)\big) \leq 2^{-\eta_i^* \gamma}, $ where $\eta_i^* > 0$ is the unique solution to $\expValDist{(f_i^a(X)/f_{k(\theta)}^a(X))^{\eta}}{f_\theta^a} = 1$.
\end{proposition}
The proof of the proposition proceeds in two steps.
First, we establish the existence of $\eta_i^*$.
Next, we apply a Chernoff-type bound to derive the stated error exponent.
\ifShowSupp
    The detailed proof is provided in Appendix \ref{prop: PHIDELTA pe bound Full proof}. 
\fi

Proposition \ref{prop: PHIDELTA pe bound Full} bounds the error probability of \PHIDELTA in a single round and shows that it decays exponentially in $\gamma$, the Armitage (or Wald) threshold required to defeat a competing hypothesis.
The result relies on the representative of the cluster containing $\theta$ being closer to $\theta$, in the \KLDtxt sense, than any other cluster representative; that is, $\DKLD{\theta}{i,}{k(\theta)}{a} > 0$ for any cluster representative $i\neq k(\theta)$.
If this result consistently holds for all iterations, we have the following:
\begin{lemma}
    Assume $\DKLD{\theta}{i,}{k(\theta)}{a} > 0$ for each iteration of \PHIDELTA.
    Let $\eta^\star$ be the smallest of $\eta_i^*$ among iterations and competitors in each iteration.
    Then, $p_e(\text{\PHIDELTA})\leq (H-1)\times2^{-\eta^\star\gamma}$.
\end{lemma}
The proof of this lemma follows by bounding the error bound in Proposition \ref{prop: PHIDELTA pe bound Full} by $2^{-\eta^\star\gamma}$, and proceeding to apply the union bound twice as in the proof of Lemma \ref{lemma: Phi pe bound}.

The bound in Proposition \ref{prop: PHIDELTA pe bound Full} depends on $\eta^\star$, which may be smaller than one.
In such cases, \PHIDELTA may exhibit a slower error-decay rate than $\Phi$.
To address this limitation, we shift our focus to the \SEF.
In this setting, we first show that the clustering procedure introduced in Subsection \ref{subsection: clustering step} guarantees the cluster-representative property required by Proposition \ref{prop: PHIDELTA pe bound Full}.
Moreover, we prove that, for the \SEF, the error bound of \PHIDELTA matches that of $\Phi$.

\begin{lemma}
    \label{lemma: exp family clustering properties}
    If $\{f_h^a\}_{h\in\mathcal{H}}$ are from the \emph{same} \SEF, then:
    \begin{enumerate}
        \item If $T(x)$ is monotone, then the \TVDtxt-induced clusters are contiguous.
        That is, if $i, j \in \equivClusterIAEps{\mathcal{U}}$, then $l\in\equivClusterIAEps{\mathcal{U}}$ for any $l\in\mathcal{U}$ such that $\eta_l\in[\min\{\eta_i,\eta_j\}, \max\{\eta_i,\eta_j\}]$. 
        \item If $k(\theta) = \argmax{ \eta_l }{l\in\equivClusterThetaAEps{\mathcal{U}}}$, then $\DKLD{\theta}{i,}{k(\theta)}{a} > 0$ for any $i$ such that $\eta_{k(\theta)} < \eta_i$.
        Particularly, $\expValDist{ f_{i}^{a}(X) / f_{k(\theta)}^{a}(X) }{ f_\theta^{a} } \leq 1$ decreases as $\eta_i$ increases.
        \item If $k(\theta) = \argmin{ \eta_l }{l\in\equivCluster{\theta}{\mathcal{U}}{a}{\varepsilon_a}}$, then $\DKLD{\theta}{i,}{k(\theta)}{a} > 0$ for any $i$ such that $\eta_i < \eta_{k(\theta)}$.
        Particularly, $\expValDist{ f_{i}^{a}(X) / f_{k(\theta)}^{a}(X) }{ f_\theta^{a} } \leq 1$ decreases as $\eta_i$ decreases.
    \end{enumerate}
\end{lemma}
\begin{IEEEproof}
    \ifShowSupp
        See Appendix \ref{lemma: exp family clustering properties proof}.
    \fi    
\end{IEEEproof}
Notably, 1) implies that a distribution from one cluster is never chosen to represent another.
2) and 3) imply that the \DM is less likely to confuse $\hTheta$ with hypotheses outside the clusters adjacent to its own.
Furthermore, 2) and 3) establish that cluster representatives can be selected to be the closest to the cluster boundaries, thereby devising a representative selection rule that is independent of the realized value of $\theta$. 

\begin{lemma}
    \label{lemma: PHIDELTA pe bound}
    Fix $a\in\mathcal{A}$ and $\mathcal{U}\subseteq \mathcal{H}$.
    Set $L_{ij} = 0$ for each $i, j\in\mathcal{U}$ before conducting the Armitage Test.
    Assume that $\{f_h^a\}_h$ are from the same \SEF.
    If $k(\theta)$ is selected as in Lemma \ref{lemma: exp family clustering properties} competes against $i$, then the per-iteration error probability does not exceed $2^{-\gamma}$.
    Particularly, $p_e(\text{\PHIDELTA})\leq (H-1)\times2^{-\gamma}$.
\end{lemma}
\begin{IEEEproof}
    \ifShowSupp
        See Appendix \ref{lemma: PHIDELTA pe bound proof}.
    \fi
    %
\end{IEEEproof}

We now turn to the derivation of an error-probability bound for the $I$ algorithm.
\begin{lemma}
    \label{lemma: IOTA pe bound}
    $p_e(I)\leq (H-1)\times 2^{-\gamma}$.
\end{lemma}
\begin{IEEEproof}
    With a similar computation made by Chernoff in \cite{Chernoff1959SequentialHT}, we bound the conditional (on the stopping time) per-iteration error probability upon iteration termination by $2^{-\gamma}$.
    Thus, the non-conditional probability to err does not exceed $2^{-\gamma}$.
    Since there are at most $H-1$ other hypotheses when erring, we obtain an additional $H-1$ factor from the union bound.
    \ifShowSupp
        For details, see Appendix \ref{lemma: IOTA pe bound proof}.
    \fi
\end{IEEEproof}

The upper bounds on the error probabilities of all three algorithms, $\Phi$, \PHIDELTA, and $I$, depend on the competition stopping threshold $\gamma$.
This relationship can be inverted: for any prescribed maximal error probability, one can choose $\gamma$ accordingly.
The following theorem summarizes this implication for the three algorithms.
\begin{theorem}[Error Probability’s Scaling Laws in $\delta$]
    \label{theorem: pe bound}
    For any $\delta \in (0,1)$, there exists a choice of the stopping threshold $\gamma$ such that $p_e(\Phi), p_e(\text{\PHIDELTA}), p_e(I) \leq \delta$.
\end{theorem}
The result follows directly by substituting $\gamma = \log((H-1)/\delta)$ into the error bounds established in Lemma \ref{lemma: Phi pe bound}, Lemma \ref{lemma: PHIDELTA pe bound}, and Lemma \ref{lemma: IOTA pe bound} for $\Phi$, \PHIDELTA, and $I$, respectively.

The error bounds derived hold in general, but can be sharpened when the induced distributions under the selected action belong to the same \SEF.
Specifically, Lemma \ref{lemma: exp family clustering properties} enables a more aggressive clustering scheme that accounts not only for the inter-cluster distance between hypotheses, but also for the number of hypotheses assigned to each cluster.
In particular, one may partition distributions into two roughly equal-sized clusters in the parameter space, ensuring that the two clusters are \TVDtxt-separated by at least $\varepsilon_a$.
This partition will allow the \DM to prune roughly half of the hypotheses in a single iteration, making the elimination framework as effective as binary search.
Consequently, the total error bound improves from $\bigO{H\times2^{-\gamma}}$ to $\bigO{(\log H)\times2^{-\gamma}}$, significantly improving over the classical search-based schemes, e.g., \cite{Armitage1950_SHT_MultipleHypotheses, Chernoff1959SequentialHT, Cohen_Zhao2015_SHT_AnomalyDetection}.
Moreover, this partition allows taking smaller \LLR thresholds, e.g., $\gamma=\bigO{\log((\log H)/\delta)}$ rather than $\bigO{\log(H/\delta)}$, which in turn implies that fewer samples can be taken as we argue in the following subsection.

\subsection{Bounding the Expected Number of Samples}
\label{subsection: Na Analysis}
Denote by $N_a^\Gamma\in\mathbb{N}$ the number of times action $a\in\mathcal{A}$ is applied under $\Gamma\in\{\Phi, \text{\PHIDELTA}, I\}$.
We start by analyzing the expected number of samples used by $\Phi$.
Since $\Phi$ relies on successive Armitage Tests, the per-iteration number of samples is given by \cite[Theorem~4.1]{Dragalin_etAl_1999_MSPRT_AsympOpt}:
\begin{lemma}
    \label{lemma: per-iteration Na Phi}
    Fix $a\in\mathcal{A}$ and $\mathcal{U}\subseteq \mathcal{H}$.
    Then, 
    $\expVal{\NaP | \theta} = (1+\littleO{1})\gamma / \min_{ j\not\in \equivClass{\theta}{\mathcal{U}}{a} } \KLDthetai$, where the little-O term is with respect to $\gamma\to\infty$.
\end{lemma}

The expected number of samples under \PHIDELTA is similar to $\Phi$, except being governed by the \KLDtxt difference in Eq. \eqref{eq: KLD Difference}.
Formally:
\begin{lemma}
    \label{lemma: per-iteration Na PHIDELTA}
    Fix $a\in\mathcal{A}$ and $\mathcal{U}\subseteq \mathcal{H}$.
    Set $L_{ij} = 0$ for each $i, j\in\mathcal{U}$ before conducting the Armitage Test.
    Let $k(\theta)$ denote the hypothesis representing $\equivClusterThetaAEps{\mathcal{U}}$, and $\mathcal{I}$ be the set of its contestants.
    Then, if $\DKLD{\theta}{i,}{k(\theta)}{a} > 0$ for any $i\in\mathcal{I}$, then $\expVal{\NaPD | \theta} = (1+\littleO{1})\gamma / \min_{i\in\mathcal{I}} \DKLD{\theta}{i,}{k(\theta)}{a}$.
\end{lemma}
\begin{IEEEproof}
    Leveraging Assumption \ref{assumption: finite LLR variance} and Kolmogorov’s Maximal Inequality \cite[Theorem~22.4]{Billingsley1995_Prob_and_Measure}, the accumulated \LLRs concentrate around their means.
    Thus, since $\DKLD{\theta}{i,}{k(\theta)}{a} > 0$, we can invoke the Strong Law of Large Numbers in \cite[Theorem~3.1]{Dragalin_etAl_1999_MSPRT_AsympOpt} to obtain the result.
    \ifShowSupp
        For details, see Appendix \ref{lemma: per-iteration Na PHIDELTA proof}.
    \fi    
\end{IEEEproof}

Bounding the total number of samples in the $I$ algorithm is more challenging than in $\Phi$ or \PHIDELTA for two reasons.
The first is the lack of structure in the order of pruned hypotheses when conditioning on $\theta$.
Namely, the condition under which $I$ stops applying a specific action and starts applying a new one (i.e., action switching times) does not depend on the number of winners or losers, unlike \PHIDELTA, which switches actions after identifying a single iteration winner.
The second challenge stems from the fact that switching from one action to another does not prevent $I$ from re-selecting it in a later iteration.
This behavior contrasts with \PHIDELTA, which never re-selects actions.

The dependence on the elimination order complicates the ability to derive meaningful bounds on $\expVal{N|\theta, \Gamma = I}$, and, accordingly, we must devise bounds that are independent of the elimination order to some degree.
To this end, we define a new probability measure on the actions, $\lambda_i^a|\Gamma$, that assigns a non-zero probability to action $a$ that is proportional to its usage when $\hII$ is true and zero otherwise.
That is, $\lambda_i^a|\Gamma \triangleq \expVal{N_a^{\Gamma}|\theta = i} / \expVal{N|\theta = i, \Gamma}$.
For any sequence indexed by actions $\{b_a\}_{a\in\mathcal{A}}$, we write $\sum_{a\in\mathcal{A}} b_a \times\lambda_i^a|\Gamma \triangleq \expValDist{b_{A}|\Gamma}{A\sim\lambda_i}$ to simplify notation.
The following lemma provides general bounds on $\expVal{N|\theta, \Gamma}$:
\begin{lemma}
    \label{lemma: N bound general}
    Let $D_j^{\Gamma}\triangleq \expValDist{ \KLD{ f_\theta^A }{ f_j^A } | \Gamma  }{A\sim\lambda_\theta}$.
    Any Wald Test-based procedure $\Gamma$ with $p_e(\Gamma)\leq 2^{-\gamma}$ has an upper bound of the form $\expVal{N|\theta, \Gamma} = (1+\littleO{1}) \gamma / \min_{j\neq \theta} D_j^{\Gamma}$ and an asymptotically matching lower bound of the form $\expVal{N|\theta, \Gamma} = (1-\littleO{1}) \gamma / \min_{j\neq \theta} D_j^{\Gamma}$.
\end{lemma}
\begin{IEEEproof}
    The proof follows by decomposing the expected accumulated \LLR between $\theta$ and $j$ into $\expVal{N|\theta, \Gamma}\times D_j$.
    We then leverage the fact that upon termination, $\abs{L_{\theta j}} = \gamma + \sigma_{\theta j}(a_N)$, where $\sigma_{\theta j}(a_N)\geq 0$ is a random variable quantifying the overshoot from $\gamma$ upon termination, to upper bound $\expVal{\abs{L_{\theta j}}}$ using Lorden’s Inequality \cite{lorden1970excess}.
    The bound on $\expVal{\abs{L_{\theta j}}}$ is then utilized to bound $\expVal{L_{\theta j}}$ from both sides.
    \ifShowSupp
        For details, see Appendix \ref{lemma: N bound general proof}.
    \fi    
\end{IEEEproof}
Note that we can mitigate the effects of the above-mentioned challenges by weakening the bound by noticing that, for any $j\neq\theta$, the mean \KLDtxt is bounded by: 
\begin{align}
    \label{eq: mean KLD upper and lower bounds}
    \min_{ \substack{ a\in\mathcal{A} \\ i \not\in \equivClass{\theta}{\mathcal{H}}{a} } } \hspace{-16pt} \KLD{ f_\theta^a }{ f_i^a } 
    \leq
    D_j^{\Gamma}
    \leq
    \max_{ \substack{i\neq \theta \in\mathcal{H}\\a\in\mathcal{A}} } \KLD{ f_\theta^a }{ f_i^a }
    .
\end{align}

We conclude this subsection by summarizing the scaling laws of the number of samples in $\delta$ in the following:
\begin{theorem}[Expected Number of Samples’ Scaling Laws in $\delta$]
    \label{theorem: Na bound}
    For any $\delta \in (0,1)$, there exists a choice of the stopping threshold $\gamma$ such that $\expVal{N | \Gamma = \Phi}$, $\expVal{N | \Gamma = \text{\PHIDELTA}}$, $\expVal{N | \Gamma = I} = \bigTheta{\log(1/\delta)}$.
\end{theorem}
Similar to Theorem \ref{theorem: pe bound}, this result follows by substituting $\gamma = \log((H-1)/\delta)$ into the Lemmas \ref{lemma: per-iteration Na Phi}, \ref{lemma: per-iteration Na PHIDELTA}, and \ref{lemma: N bound general}.

\subsection{Asymptotic Optimality and Performance Improvement in the Finite Regime}
\label{subsection: optimality}
We first establish asymptotic optimality in $\delta\to0$ by combining Theorems \ref{theorem: pe bound} and \ref{theorem: Na bound}:
\begin{corollary}[Asymptotic Optimality in $\delta$]
    \label{corollary: vanishing ABR}
    Setting $\gamma = \log ((H-1)/\delta)$ in either $\Phi$, \PHIDELTA, or $I$ results in $\eqref{eq: Bayes Risk} = \bigTheta{\delta \log(1/\delta)}$
    Particularly, the \ABR vanishes as $\delta\to 0$.
\end{corollary}
\begin{IEEEproof}
    Combining the form in Lemma \ref{lemma: N bound general} with Theorems \ref{theorem: pe bound} and \ref{theorem: Na bound}, for any fixed $H$ there are $\delta_0$, $c_1$, $c_2 > 0$ such that $c_1 \log (1/\delta) \leq \expVal{N | \Gamma}\leq c_2\log(1/\delta)$ for any $\delta < \delta_0$.
    Accordingly, $c_1 \delta\log (1/\delta) \leq  \eqref{eq: Bayes Risk} \leq c_2\delta\log(1/\delta) + \delta$ for any $\delta < \delta_0$.
    Notably, $\delta \log(1/\delta)\to 0$ as $\delta\to 0$.
\end{IEEEproof}

We now shift our focus to improve the sample complexity in the finite regime.
Since all of our proposed algorithms use deterministic action-selection rules, the action sequences used to recover $\theta$ successfully can either be computed in advance for any $\theta$ (in $\Phi$ or \PHIDELTA) or are predictable (in $I$, in the sense that the hypothesis pruning process in $I$ admits only finitely many relevant permutations until an action is swapped).
Accordingly, $\expVal{N|\theta, \Gamma}$ can be computed in advance, enabling one to optimize the conditional sample complexity.
However, we argue that this optimization can be computationally prohibitive:
\begin{theorem}
    \label{theorem: opt action seq is NP-Hard}
    Determining the optimal action selection minimizing $\expVal{N|\Gamma}$ for multi-iteration \SHT algorithms is NP-hard.
\end{theorem}
\begin{IEEEproof}
    We reduce the multi-iteration \SHT to the \MWDT problem \cite{garey_johnson2002computers}.
    \ifShowSupp
        For details, see Appendix \ref{theorem: opt action seq is NP-Hard proof}.
    \fi        
\end{IEEEproof}
In other words, Theorem \ref{theorem: opt action seq is NP-Hard} establishes that $\Phi$ and $I$ are optimal in the sense that for the correct action sequences, the minimal expected number of samples required to identify the true hypothesis can be achieved (with $I$ outperforming $\Phi$ due to its finer granularity).
This result naturally holds for \PHIDELTA when $\{\varepsilon_a\}_{a}$ \emph{are fixed} and the cluster representatives are known.
Namely, \PHIDELTA is even more complex since the selection of $\{\varepsilon_a\}_{a}$ and cluster representatives must also be considered.

\subsection{Space and Average Runtime Complexity}
\label{subsection: Space and Average Runtime Complexity}
In this subsection, we analyze the complexity of our algorithms, starting with space complexity.
The space complexity during runtime is dominated by the need to track $\bigO{H^2}$ \LLRs, so the space complexity of $\Phi$ and $I$ is $\bigO{H^2}$.
For \PHIDELTA, saving the clustering results requires additional $\bigO{\abs{\mathcal{A}}H}$ space, totaling in a space complexity of $\bigO{\abs{\mathcal{A}}H + H^2}$.

We shift our focus to average runtime.
Consider a single iteration.
To compute the action, our algorithms compare the \TVDtxt between all alive hypotheses, taking $\bigO{\abs{\mathcal{A}} H^2}$ steps.
Once the action is computed, computing the contestants takes $\bigO{H}$ steps.
The sample acquisition and the pairwise \LLR update until iteration termination take $\bigO{ \expVal{N_a^\Gamma|\theta} + H^2 }$ steps on average.
Updating the alive hypotheses requires additional $\bigO{H}$ steps.
Accordingly, since there are at most $H-1$ iterations, the total average runtime is $\bigO{H\times(\abs{\mathcal{A}} H^2 + H + \expVal{N_a^\Gamma|\theta} + H^2 + H)} = \bigO{\abs{\mathcal{A}} H^3 + H\expVal{N_a^\Gamma|\theta}}$.

\section{Numerical Results}
\label{section: Numerical Results}
In this section, we present simulation results comparing the performance of our algorithms with other algorithms that offer strong theoretical guarantees.
Specifically, we compare our algorithms with \cite[Algorithm~2]{Gan_Jia_Li2021_Decision_Tree_SHT}, which will be dubbed as GJL, the seminal Chernoff scheme \cite{Chernoff1959SequentialHT}, and against Policy 1 \cite[Section~4.2]{Naghshvar_Javidi2013_SHT_DynamicProgramming}, which will be dubbed as NJ1\footnote{We omit numerical comparisons with Policy 2 in \cite{Naghshvar_Javidi2013_SHT_DynamicProgramming} since it requires an intractable computation.
In particular, it requires optimizing the \KLDtxt separation between each distribution and its corresponding \emph{optimal} mixture of the others. This \KLDtxt admits no closed-form solution due to the weighted sum inside the logarithm, even for simple cases (e.g., \cite{Hershey_Olsen2007Approximating_GMM_KLD}).}.

GJL is an adaptive multi-stage deterministic algorithm, in which the \DM selects the action that eliminates the maximum number of hypotheses while minimizing the number of overlapping hypotheses in terms of the output distribution’s mean.
That is, the sample output distribution under each action must be limited to single-parameter distributions that can be identified by computing the empirical mean (e.g., Bernoulli, exponential, and unit-variance normal distributions).
After some action is selected, a \emph{fixed} number of samples is collected, where the number is determined by the number of samples required to separate the closest alive yet non-identical hypotheses under the selected action.
This process repeats until only one hypothesis remains.

The Chernoff scheme is a simple, asymptotically optimal, adaptive stochastic algorithm in which, unlike in our model, all actions are assumed to be able to separate all hypotheses\footnote{the Chernoff scheme retains its asymptotic optimality in our model, as argued in \cite[Theorem~2]{Cohen_Zhao2015_SHT_AnomalyDetection} and \cite{Naghshvar_Javidi2013_SHT_DynamicProgramming}.}.
At each time step, an action is drawn at random and applied to obtain a sample, which is used to update the posterior probabilities of the hypotheses until some hypothesis has a posterior probability greater than $1-\delta$, at which point it is declared true.
The distribution from which actions are drawn is constructed to favor separating the currently most likely hypothesis from the others.

NJ1 relaxes Chernoff’s separation assumption to be Assumption \ref{assumption: validity}, and adds an exploration phase before using the Chernoff scheme.
In the exploration phase, which persists as long as no hypothesis has its posterior probability above some tunable confidence $\tilde{\rho} > 0.5$, actions are drawn according to their worst-case separation.

We conduct two simulation scenarios, each constructed so that \PHIDELTA must use at least two iterations.
To give NJ1 and Chernoff an edge over our proposed schemes, the \LLR threshold $\gamma$ is fixed to be $\log((H-1)/\delta)$ (i.e., the hypothesis structure is not leveraged beyond clustering).
The proximity parameters $\{\varepsilon_a\}$ in \PHIDELTA were selected to minimize the Dunn Index \cite[Chapter~23]{aggarwal2014data}, and remain fixed in each simulation scenario.
For a fair comparison with GJL, the distributions from which samples are drawn can be identified solely by their empirical means.
The simulations conducted are motivated by classic wireless signal detection (e.g. \cite[Chapter~2.2.4]{tse_2005}, \cite[Chapter~5]{goldsmith2005}, \cite[Chapter~4]{Proakis2007}) in multi-user settings.

We consider a single receiver, acting as the \DM, whose goal is to identify the active transmitter among $H=32$ possible users.
The active user transmits a unique identifier, represented as a binary sequence, to the receiver across the available frequency bands.
Accordingly, the receiver acquires information by sequentially probing frequency bands, where each probe corresponds to an action available to the \DM.

The statistical behavior of the samples depends on both the active user and the selected frequency band.
Specifically, each frequency band is affected by fading and additive noise, which jointly determine the received \SNR.
The fading component is band-dependent and user-specific, thereby inducing distinct sample distributions for each user-band pair.
In each simulation scenario, the fading coefficients are drawn once at the beginning and then kept fixed throughout the simulation.
In contrast, the noise is resampled independently at each probe and is modeled as a Gaussian random variable.
Thus, while the fading coefficients define the underlying hypothesis-dependent structure of the problem, the Gaussian noise generates the sample-to-sample variability observed by the \DM.

In the first scenario, on each band (out of $\abs{\mathcal{A}} = 16$), the active user transmits either $-8$ or $8$, corresponding to the binary symbols ‘0’ or ‘1’, respectively.
The received signal is then affected by a user- and band-dependent fading term, modeled as an additive random offset drawn uniformly from $[-1,1]$, together with normalized additive white Gaussian noise.
Consequently, for any fixed user and probed frequency band, the receiver observes Gaussian samples whose mean is determined by the transmitted symbol and the corresponding fading realization with unit variance.

To make the identification task more challenging, we construct two users who are nearly indistinguishable from each other.
In particular, users 0 and 31 are assigned the same identifier and the same fading coefficients on all bands except the last one.
This setting can be interpreted as two applications or services operating over the same physical transmitter, differing only in a single identifying bit.
As a result, when the true hypothesis is $\theta\in\{0, 31\}$, all actions except probing the last band induce identical sample distributions under the two hypotheses.
Therefore, distinguishing between these two users requires probing the unique frequency band on which their identifiers differ.

\begin{figure}[!htbp]
    \centering
    \includegraphics[scale=0.7]{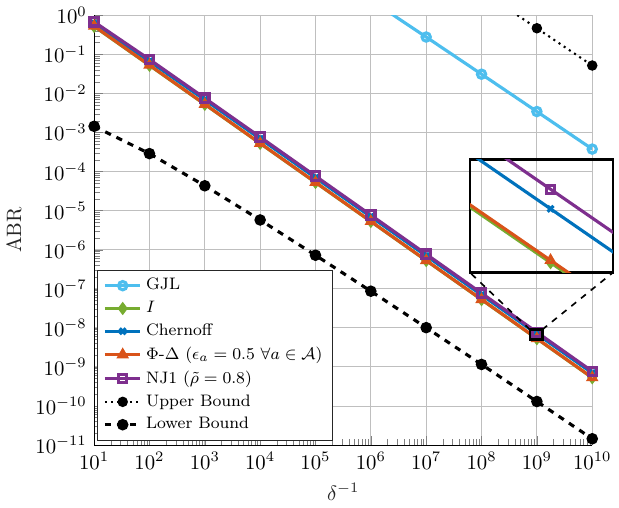}
    \vspace{-8pt}
    \caption{
        The \ABR in the first wireless network scenario.
        Here, we have $H = 32$, $\abs{\mathcal{A}} = 16$, and all samples follow unit-variance normal distributions with randomly drawn fixed means.
    }
    \label{figure: avg Bayes Risk vs. delta eps norm}
\end{figure}

The simulation results of the first scenario are given in Figure \ref{figure: avg Bayes Risk vs. delta eps norm}, where the log-log plot of the empirical \ABR as a function of $1/\delta$ is presented.
The purple, blue, orange, green, and cyan curves correspond to NJ1 (with $\tilde{\rho} = 0.8$), the Chernoff scheme, \PHIDELTA (with $\varepsilon_a = 0.5$ for any $a$), $I$, and GJL, respectively.
The \ABR of $I$ and \PHIDELTA almost coincide in this simulation, with $I$ outperforming \PHIDELTA (attributed to early action switch).
The black-dashed and black-dotted lines are the lower and upper bounds in Eq. \eqref{eq: mean KLD upper and lower bounds} substituted into Lemma \ref{lemma: N bound general}. 

Overall, the simulation results highlight two key points discussed in this paper.
First, the proposed elimination-based strategies preserve the desired $\log(1/\delta)$ scaling of the sample complexity.
This behavior is reflected in the eventual linear decay of the empirical \ABR on $\log(1/\delta)$.
Second, the results demonstrate the importance of exploiting the underlying structure among the hypotheses, rather than treating the hypothesis set as unstructured.

This distinction is apparent from the empirical \ABR values.
Since the sample distributions differ only through small random offsets, the worst-case separation between some hypothesis pairs is relatively low.
Consequently, the structure-oblivious GJL algorithm is forced to account for the most difficult pairwise distinctions and is therefore significantly outperformed by the other algorithms.
In contrast, \PHIDELTA explicitly clusters similar distributions, leading in this setting to two dominant clusters under each action, whereas $I$ inherently adapts to the distributional structure of the hypotheses.
These structural advantages allow both algorithms to achieve the best empirical performance.
Moreover, \PHIDELTA eliminates a fraction close to half of the remaining hypotheses at each iteration, similarly to the aggressive clustering scheme discussed in Section \ref{subsection: PE Analysis}.

In the second scenario, we consider a non-coherent reception in which the receiver employs energy detectors \cite[Chapter~3.1.1]{tse_2005} rather than conventional symbol detectors.
Under this scheme, each bit of the identifier is conveyed through on-off transmission over one of $\abs{\mathcal{A}} = 10$ frequency bands: transmitting a ‘1’ corresponds to sending a signal over the selected band, whereas transmitting a ‘0’ corresponds to remaining silent.
Thus, probing a band amounts to measuring the received energy on that band and using it to infer whether the active user transmitted on it.

The channel in this simulation follows a Rayleigh fading model at the high \SNR regime.
The noise power is set to 10, while the transmission power is 5000.
The fading effects introduce mild user- and band-dependent variability drawn uniformly from $[-5,5]$ (i.e., the \SNR is approximately 27dB).
Consequently, the receiver observes exponentially distributed energy samples: when the active user is silent on the probed band, the samples have a mean of 10, whereas when the active user transmits, the mean is dominated by the transmission power and is therefore perturbed around 5000.
Hence, the resulting sample distributions are highly separated, reflecting the strong contrast between silent and active bands.

\begin{figure}[!htbp]
    \centering
    \includegraphics[scale=0.7]{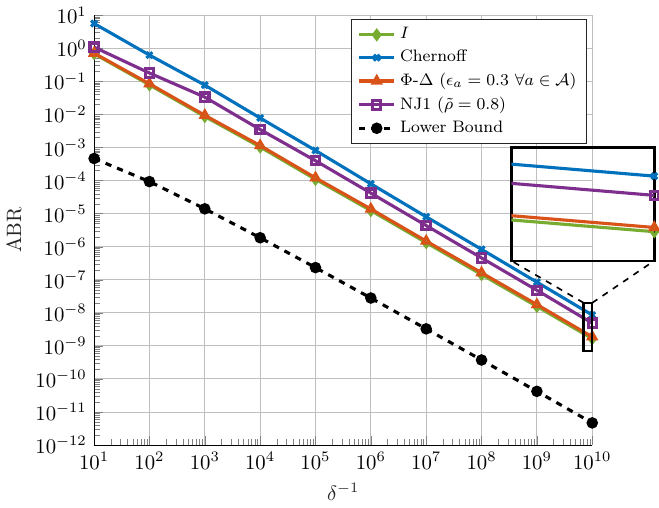}
    \vspace{-8pt}
    \caption{
        The \ABR in the second wireless network simulation.
        Here, we have $H = 32$, $\abs{\mathcal{A}} = 10$, and all samples follow exponential distributions.
    }
    \label{figure: avg Bayes Risk vs. delta eps exp}
\end{figure}

The results of the second scenario can be found in Figure \ref{figure: avg Bayes Risk vs. delta eps exp}.
The figure compares \PHIDELTA, with $\varepsilon_a = 0.3$ for any $a$, and $I$ against NJ1, the Chernoff scheme, and GJL.
As in the previous simulation, both \PHIDELTA and I consistently outperform the competing schemes.
In particular, $I$ requires roughly an order of magnitude fewer samples than the Chernoff scheme for the same prescribed error probability.

Unlike the previous figure, the \ABR curve of GJL and the corresponding upper bounds are omitted from Figure \ref{figure: avg Bayes Risk vs. delta eps exp}.
This is because the lowest \ABR attained by GJL in this setting is approximately 180, which is also close to the scale of the upper bounds; including these curves would therefore obscure the behavior of the remaining algorithms in the relevant range of the plot.
Finally, as in the first scenario, the elimination framework remains effective: approximately half of the hypotheses are pruned at each iteration of \PHIDELTA.

\section{Conclusions}
In this work, we studied active \SHT through a multi-iteration elimination framework.
We proposed the $\Phi$, \PHIDELTA, and $I$ algorithms and analyzed their error probabilities, sample complexities, computational complexities, and asymptotic behavior.
We showed that elimination-based designs can provide as strong theoretical guarantees as their traditional search counterparts while offering a structured, computationally interpretable approach to sequential decision-making.
Particularly, \PHIDELTA demonstrates that clustering hypotheses can accelerate elimination, while $I$ highlights the benefit of targeting the rapid removal of “almost surely” incorrect hypotheses.
Note that hypothesis clustering can be made more aggressive when samples are drawn from the \SEF distributions, making the elimination framework as effective as binary search.

Several directions remain open, including: (i) improved clustering mechanisms, (ii) selecting cluster representatives beyond the \SEF, (iii) studying multidimensional distributions under the hypothesis-clustering framework, and (iv) constructing virtual cluster representatives.

\ifShowAppendix
    \appendices
\fi

\bibliographystyle{IEEEtran}
\bibliography{references}

@ARTICLE{Nitinawarat2013_SHT_Argmax2_KLD_wProofs,
  author={Nitinawarat, Sirin and Atia, George K. and Veeravalli, Venugopal V.},
  journal={IEEE Transactions on Automatic Control}, 
  title={Controlled Sensing for Multihypothesis Testing}, 
  year={2013},
  volume={58},
  number={10},
  pages={2451-2464},
  doi={10.1109/TAC.2013.2261188}
}

@article{Naghshvar_Javidi2013_SHT_DynamicProgramming,
  author = {Naghshvar, Mohammad and Javidi, Tara},
  journal = {The Annals of Statistics},
  title = {Active Sequential Hypothesis Testing},
  year = {2013},
  volume = {41},
  number={6},
  pages = {2703-2738},
  doi = {10.1214/13-AOS1144}
}

@ARTICLE{Cohen_Zhao2015_SHT_AnomalyDetection,
  author={Cohen, Kobi and Zhao, Qing},
  journal={IEEE Transactions on Information Theory}, 
  title={Active Hypothesis Testing for Anomaly Detection}, 
  year={2015},
  volume={61},
  number={3},
  pages={1432-1450},
  doi={10.1109/TIT.2014.2387857}
}

@book{CoverThomas2006,
    author = {Cover, Thomas M. and Thomas, Joy A.},
    title = {Elements of Information Theory (Wiley Series in Telecommunications and Signal Processing)},
    year = {2006},
    isbn = {0471241954},
    publisher = {Wiley-Interscience},
    address = {USA}
}

@book{Cormen2009IntroToAlgo3,
  author       = {Thomas H. Cormen and
                  Charles E. Leiserson and
                  Ronald L. Rivest and
                  Clifford Stein},
  title        = {Introduction to Algorithms, 3rd Edition},
  publisher    = {{MIT} Press},
  year         = {2009},
  url          = {http://mitpress.mit.edu/books/introduction-algorithms},
  isbn         = {978-0-262-03384-8},
  bibsource    = {dblp computer science bibliography, https://dblp.org}
}

@article{Chernoff1959SequentialHT,
    author = {Herman Chernoff},
    title = {Sequential Design of Experiments},
    volume = {30},
    journal = {The Annals of Mathematical Statistics},
    number = {3},
    publisher = {Institute of Mathematical Statistics},
    pages = {755 -- 770},
    year = {1959},
    doi = {10.1214/aoms/1177706205},
    URL = {https://doi.org/10.1214/aoms/1177706205}
}

@ARTICLE{Bai_Katewa_Gupta_Huang2015_Stochastic_Source_Selection,
  author={Bai, Cheng-Zong and Katewa, Vaibhav and Gupta, Vijay and Huang, Yih-Fang},
  journal={IEEE Transactions on Signal Processing}, 
  title={A Stochastic Sensor Selection Scheme for Sequential Hypothesis Testing With Multiple Sensors}, 
  year={2015},
  volume={63},
  number={14},
  pages={3687-3699},
  doi={10.1109/TSP.2015.2425804}
}

@ARTICLE{Bar_Tabrikian2018_Composite_SHT_Single_Source,
  author={Bar, Shahar and Tabrikian, Joseph},
  journal={IEEE Transactions on Signal Processing}, 
  title={A Sequential Framework for Composite Hypothesis Testing}, 
  year={2018},
  volume={66},
  number={20},
  pages={5484-5499},
  doi={10.1109/TSP.2018.2866813}
}

@article{Wald_1945_SHT,
 ISSN = {00034851},
 URL = {http://www.jstor.org/stable/2235829},
 author = {A. Wald},
 journal = {The Annals of Mathematical Statistics},
 number = {2},
 pages = {117--186},
 publisher = {Institute of Mathematical Statistics},
 title = {Sequential Tests of Statistical Hypotheses},
 urldate = {2024-06-13},
 volume = {16},
 year = {1945}
}

@article{Wald_Wolfowitz_1948_SPRT_Optimality,
 ISSN = {00034851},
 URL = {http://www.jstor.org/stable/2235638},
 author = {A. Wald and J. Wolfowitz},
 journal = {The Annals of Mathematical Statistics},
 number = {3},
 pages = {326--339},
 publisher = {Institute of Mathematical Statistics},
 title = {Optimum Character of the Sequential Probability Ratio Test},
 urldate = {2026-05-14},
 volume = {19},
 year = {1948}
}

@inproceedings{Gan_Jia_Li2021_Decision_Tree_SHT,
 author = {Gan, Kyra and Jia, Su and Li, Andrew},
 booktitle = {Advances in Neural Information Processing Systems},
 editor = {M. Ranzato and A. Beygelzimer and Y. Dauphin and P.S. Liang and J. Wortman Vaughan},
 pages = {5012--5024},
 publisher = {Curran Associates, Inc.},
 title = {Greedy Approximation Algorithms for Active Sequential Hypothesis Testing},
 url = {https://proceedings.neurips.cc/paper_files/paper/2021/file/27e9661e033a73a6ad8cefcde965c54d-Paper.pdf},
 volume = {34},
 year = {2021}
}

@article{Armitage1950_SHT_MultipleHypotheses,
 ISSN = {00359246},
 URL = {http://www.jstor.org/stable/2983839},
 author = {P. Armitage},
 journal = {Journal of the Royal Statistical Society. Series B (Methodological)},
 number = {1},
 pages = {137--144},
 publisher = {[Royal Statistical Society, Wiley]},
 title = {Sequential Analysis with More than Two Alternative Hypotheses, and its Relation to Discriminant Function Analysis},
 urldate = {2024-08-21},
 volume = {12},
 year = {1950}
}

@INPROCEEDINGS{Citron_Cohen_Zhao2024_DGF_on_Hidden_Markov_Chains,
  author={Citron, Levli and Cohen, Kobi and Zhao, Qing},
  booktitle={{IEEE} International Symposium on Information Theory ({ISIT})}, 
  title={Anomaly Search of a Hidden Markov Model}, 
  year={2024},
  volume={},
  number={},
  pages={3684-3688},
  doi={10.1109/ISIT57864.2024.10619616}
}

@book{lehmann2006testing,
  title={Testing Statistical Hypotheses},
  author={Lehmann, E.L. and Romano, J.P.},
  isbn={9780387276052},
  lccn={2004051464},
  series={Springer Texts in Statistics},
  url={https://books.google.co.il/books?id=K6t5qn-SEp8C},
  year={2006},
  publisher={Springer New York}
}

@article{bessler1960theory,
  title={Theory and applications of the sequential design of experiments, k-actions and infinitely many experiments: Part {I}--{Theory}},
  author={Bessler, Stuart Alan},
  journal={Appl. Math. Statist. Lab., Stanford Univ., Stanford, CA, USA, Tech. Rep},
  volume={55},
  year={1960}
}

@ARTICLE{Huang2019_DGF_Heterogeneous,
  author={Huang, Boshuang and Cohen, Kobi and Zhao, Qing},
  journal={IEEE Transactions on Information Theory}, 
  title={Active Anomaly Detection in Heterogeneous Processes}, 
  year={2019},
  volume={65},
  number={4},
  pages={2284-2301},
  doi={10.1109/TIT.2018.2866257}
}

@ARTICLE{Lambez2022_DGF_wSwitchCost,
  author={Lambez, Tidhar and Cohen, Kobi},
  journal={IEEE Transactions on Signal Processing}, 
  title={Anomaly Search With Multiple Plays Under Delay and Switching Costs}, 
  year={2022},
  volume={70},
  number={},
  pages={174-189},
  doi={10.1109/TSP.2021.3136810}
}

@ARTICLE{Gurevich2019_EEST,
  author={Gurevich, Andrey and Cohen, Kobi and Zhao, Qing},
  journal={IEEE Transactions on Signal Processing}, 
  title={Sequential Anomaly Detection Under a Nonlinear System Cost}, 
  year={2019},
  volume={67},
  number={14},
  pages={3689-3703},
  doi={10.1109/TSP.2019.2918981}
}

@ARTICLE{Joseph_DeepLearining1,
  author={Joseph, Geethu and Zhong, Chen and Gursoy, M. Cenk and Velipasalar, Senem and Varshney, Pramod K.},
  journal={IEEE Sensors Journal}, 
  title={Anomaly Detection via Learning-Based Sequential Controlled Sensing}, 
  year={2024},
  volume={24},
  number={13},
  pages={21025-21037},
  doi={10.1109/JSEN.2024.3399456}
}

@ARTICLE{Szostak2024_DeepLearining2,
  author={Szostak, Hadar and Cohen, Kobi},
  journal={IEEE Access}, 
  title={Deep Multi-Agent Reinforcement Learning for Decentralized Active Hypothesis Testing}, 
  year={2024},
  volume={},
  number={},
  pages={1-1},
  doi={10.1109/ACCESS.2024.3430392}
}

@misc{stamatelis2024_DeepLearining3,
      title={Single- and Multi-Agent Private Active Sensing: A Deep Neuroevolution Approach}, 
      author={George Stamatelis and Angelos-Nikolaos Kanatas and Ioannis Asprogerakas and George C. Alexandropoulos},
      year={2024},
      eprint={2403.10112},
      archivePrefix={arXiv},
      primaryClass={cs.AI},
      url={https://arxiv.org/abs/2403.10112}, 
}

@ARTICLE{Gafni2023_CompositeHT,
  author={Gafni, Tomer and Wolff, Benjamin and Revach, Guy and Shlezinger, Nir and Cohen, Kobi},
  journal={IEEE Transactions on Signal Processing}, 
  title={Anomaly Search Over Discrete Composite Hypotheses in Hierarchical Statistical Models}, 
  year={2023},
  volume={71},
  number={},
  pages={202-217},
  doi={10.1109/TSP.2023.3242074}
}

@inproceedings{Ester_etAl_1996_DBSCAN,
author = {Ester, Martin and Kriegel, Hans-Peter and Sander, J\"{o}rg and Xu, Xiaowei},
title = {A density-based algorithm for discovering clusters in large spatial databases with noise},
year = {1996},
publisher = {AAAI Press},
booktitle = {Proceedings of the Second International Conference on Knowledge Discovery and Data Mining},
pages = {226–231},
numpages = {6},
location = {Portland, Oregon},
series = {KDD'96}
}

@ARTICLE{Dragalin_etAl_1999_MSPRT_AsympOpt,
  author={Dragalin, V.P. and Tartakovsky, A.G. and Veeravalli, V.V.},
  journal={IEEE Transactions on Information Theory}, 
  title={Multihypothesis Sequential Probability Ratio Tests .{I}. Asymptotic Optimality}, 
  year={1999},
  volume={45},
  number={7},
  pages={2448-2461},
  doi={10.1109/18.796383}
}

@ARTICLE{Dragalin_etAl_2000_MSPRT_MeanSamplesApprox,
  author={Dragalin, V.P. and Tartakovsky, A.G. and Veeravalli, V.V.},
  journal={IEEE Transactions on Information Theory}, 
  title={Multihypothesis sequential probability ratio tests. {II}. Accurate asymptotic expansions for the expected sample size}, 
  year={2000},
  volume={46},
  number={4},
  pages={1366-1383},
  doi={10.1109/18.850677}
}

@book{Aggarwal_Reddy_2013_ClusteringBook,
author = {Aggarwal, Charu C. and Reddy, Chandan K.},
title = {Data Clustering: Algorithms and Applications},
year = {2013},
isbn = {1466558210},
publisher = {Chapman \& Hall/CRC},
edition = {1st}
}

@book{garey_johnson2002computers,
author = {Garey, Michael R. and Johnson, David S.},
title = {Computers and Intractability; A Guide to the Theory of NP-Completeness},
year = {1990},
isbn = {0716710455},
publisher = {W. H. Freeman \& Co.},
address = {USA}
}

@article{lorden1970excess,
  title={On excess over the boundary},
  author={Lorden, Gary},
  journal={The Annals of Mathematical Statistics},
  volume={41},
  number={2},
  pages={520--527},
  year={1970},
  publisher={Institute of Mathematical Statistics}
}

@book{Billingsley1995_Prob_and_Measure,
  address = {New York [u.a.]},
  author = {Billingsley, {Patrick}},
  edition = {3. ed},
  isbn = {0471007102},
  pagetotal = {XII, 593},
  ppn_gvk = {164761632},
  publisher = {Wiley},
  series = {A Wiley-Interscience publication},
  title = {Probability and measure},
  url = {http://gso.gbv.de/DB=2.1/CMD?ACT=SRCHA&SRT=YOP&IKT=1016&TRM=ppn+164761632&sourceid=fbw_bibsonomy},
  year = 1995
}

@article{Hershey_Olsen2007Approximating_GMM_KLD,
  title={Approximating the Kullback Leibler Divergence Between Gaussian Mixture Models},
  author={John R. Hershey and Peder A. Olsen},
  journal={IEEE International Conference on Acoustics, Speech and Signal Processing},
  year={2007},
  volume={4},
  pages={IV-317-IV-320},
  url={https://api.semanticscholar.org/CorpusID:1290235}
}

@book{Proakis2007,
  author = {Proakis},
  publisher = {McGraw Hill},
  title = {Digital Communications 5th Edition},
  year = 2007
}

@book{tse_2005,
 author = {Tse, David and Viswanath, Pramod},
 title = {Fundamentals of Wireless Communication},
 year = {2005},
 isbn = {0521845270},
 publisher = {Cambridge University Press},
 address = {USA}
}

@book{goldsmith2005,
 place={Cambridge},
 title={Wireless Communications},
 DOI={10.1017/CBO9780511841224},
 publisher={Cambridge University Press},
 author={Goldsmith, Andrea},
 year={2005}
}

@book{aggarwal2014data,
author = {Aggarwal, Charu C. and Reddy, Chandan K.},
title = {Data Clustering: Algorithms and Applications},
year = {2013},
isbn = {1466558210},
publisher = {Chapman \& Hall/CRC},
edition = {1st}
}

\ifShowSupp
\section{Miscellaneous Proofs}
\subsection{Proof of Lemma \ref{lemma: Phi pe bound}}
\label{lemma: Phi pe bound proof}
Recall that the $\Phi$ algorithm proceeds via successive Armitage Tests between representatives of equivalence classes under the action selected at each stage, where each class consists of hypotheses inducing the same distribution under that action.

The error probability of a single Armitage Test depends on the number of competing equivalence classes $M\leq H$ and the decision threshold $\gamma$, which the \LLR of the selected hypothesis must exceed relative to all others.
Specifically, by a similar computation by Chernoff in \cite[Lemma~3]{Chernoff1959SequentialHT}, the probability of incorrectly selecting $\hII$, $i \neq \theta$, is bounded by $(M-1)\times2^{-\gamma}$.

For completeness, we provide the computation.
To prune $\theta$ after $n$ samples, the following must be satisfied:
\begin{align*}
    L_{i\theta} \geq \gamma
    \iff
    \prod_{t=1}^{n}  f_\theta^{a}(x_t) \leq 2^{-\gamma}\prod_{t=1}^{n}  f_i^{a}(x_t)
\end{align*}
Let $\mathcal{R}_{i,\theta}^{n}$ be the decision region for $\hII$ at timestep $n$ in its competition against $\hTheta$, i.e., 
\begin{align}
    \nonumber
    \mathcal{R}_{i,\theta}^{n}
    \triangleq
    \left\{
        \myVec{x}\in\mathbb{R}^{n} : \prod_{t=1}^{n}  f_\theta^{a}(x_t) \leq 2^{-\gamma} \prod_{t=1}^{n}  f_{i}^{a}(x_t)
    \right\}
\end{align}
We bound the conditional error probability:
\begin{align*}
    \nonumber
    \prob{ \text{select $i$} | \theta, n }
    &=
    \prob{ L_{i\theta} \geq \gamma | \theta, n }
    \\
    \nonumber
    &=
    \int_{\mathcal{R}_{i,\theta}^{n}} \prod_{t=1}^{n} f_{\theta}^{a}(x_t) d\indexedSet{x}{1}{n}
    \\
    &\leq
    2^{-\gamma}\int_{\mathcal{R}_{i,\theta}^{n}} \prod_{t=1}^{n} f_i^{a}(x_t) d\indexedSet{x}{1}{n}
    \\
    &\leq
    2^{-\gamma}\int_{\mathbb{R}^n} \prod_{t=1}^{n} f_i^{a}(x_t) d\indexedSet{x}{1}{n}
    =2^{-\gamma}
    .
\end{align*}
Thus, if we denote the stopping time of the iteration by $\tau_i \triangleq \inf\{ n\in\mathbb{N}: L_{i\theta} \geq \gamma \}$, we have $\prob{ \text{select $i$} | \theta} = \sum_{n=1}^{\infty} \prob{ \text{select $i$} | \theta, n } \prob{ \tau_i = n }\leq 2^{-\gamma}$.
Since exactly one of the $M$ competing hypotheses is true, the error probability is bounded by $(M-1)\times 2^{-\gamma}$ from the union bound.

Let $M_r$ denote the number of equivalence classes in iteration $r$.
For any $1\leq j \leq M_r$, denote by $m_j^r$ the number of hypotheses in each class during iteration $r$.
Assume, without loss of generality, that class 1 contains $\theta$ for all iterations.
At the beginning of each iteration, the accumulated \LLRs between all surviving hypotheses are zero; indeed, any pair of hypotheses inducing different distributions under a previously selected action would have been eliminated earlier.

The error event in iteration $r$ occurs if one of the $M_r-1$ classes not containing the true hypothesis is selected.
Since each equivalence class contains at least one hypothesis, $M_r \leq 1 + \sum_{j=2}^{M_r} m_j^r$.
Summing over all iterations yields that $\sum_{r} \sum_{j=2}^{M_r} m_j^r = H-1$ as the algorithm terminates after discarding $H-1$ hypotheses.
Thus, we obtain that
\begin{align*}
    p_e(\Phi)
    &\leq
    \sum_{r}(M_r - 1)\times2^{-\gamma}
    \leq
    \left(
        \sum_{r} \sum_{j=2}^{M_r} m_j^r
    \right) \times2^{-\gamma}
    \\
    &=
    (H-1)\times2^{-\gamma}
    .
\end{align*}

\subsection{Proof of Proposition \ref{prop: PHIDELTA pe bound Full}}
\label{prop: PHIDELTA pe bound Full proof}
We first argue the existence of $\eta_i^*$.
Let $R_{i,k(\theta)}\triangleq f_i^a(x)/f_{k(\theta)}^a(x)$ denote the likelihood ratio.
Let $\Lambda(\eta) \triangleq \expValDist{ (R_{i,k(\theta)})^\eta}{f_\theta^a}$.
Before proceeding, notice that $\Lambda(\eta)$ is at least twice differentiable (with respect to $\eta$) from Assumption \ref{assumption: finite LLR variance}.
Namely, $\Lambda(\eta)$ is the \MGF of $\log (f_i^a(X)/f_{k(\theta)}^a(X))$, and its second moment is bounded by
\begin{align*}
    &\expValDist{ \left(\log \frac{f_\theta^a(X)}{f_j^a(X)} - \log \frac{f_\theta^a(X)}{f_i^a(X)}\right)^2 }{f_\theta^a}
    \\
    &\quad\quad\quad\quad\leq
    2\Xi - 2\expValDist{ \left(\log \frac{f_\theta^a(X)}{f_j^a(X)}\right)\times \left(\log \frac{f_\theta^a(X)}{f_i^a(X)}\right) }{f_\theta^a}
\end{align*}
which is finite.
To see this, apply the Cauchy-Schwarz Inequality: $\left|\expValDist{ (\log \frac{f_\theta^a(X)}{f_j^a(X)})\times (\log \frac{f_\theta^a(X)}{f_i^a(X)}) }{f_\theta^a}\right|\leq \Xi < \infty$.

Differentiating $\Lambda(\eta)$ with respect to $\eta$ results in
\begin{align*}
    \partialDerive{\eta}{\Lambda}
    =
    \expValDist{ (R_{i,k(\theta)})^{\eta} \ln R_{i,k(\theta)} }{f_\theta^a}
    .
\end{align*}
Accordingly, $\partialDerive{\eta}{\Lambda}(0) = \expValDist{ \ln R_{i,k(\theta)}}{f_\theta^a} < 0$.
Thus, from continuity, there exists some $\tilde{\eta}$ such that $\Lambda(\eta) < 1$ for any $\eta\leq \tilde{\eta}$.
Now, we bound $\Lambda(\eta)$ from below.
Fix $\epsilon>0$.
\begin{align*}
    \Lambda(\eta)
    &\geq
    \expValDist{ \left(R_{i,k(\theta)}\right)^{\eta}\indicator{ R_{i,k(\theta)} \geq 1 + \epsilon } }{f_\theta^a}
    \\
    &\geq
    (1+\epsilon)^{\eta} \expValDist{\indicator{ R_{i,k(\theta)} \geq 1 + \epsilon } }{f_\theta^a}
    \\
    &=
    (1+\epsilon)^{\eta} \prob{ R_{i,k(\theta)} \geq 1 + \epsilon \ \middle| X\sim f_\theta^a }
\end{align*}
Recall that $\{f_h^a\}_h$ share the same support from Assumption \ref{assumption: finite LLR variance}, i.e., $f_{\theta}^a > 0$ implies that $f_i^a$, $f_{k(\theta)}^a > 0$.
Since $\TVD{f_i^a-f_{k(\theta)}^a} > \varepsilon_a$, $f_i^a \neq f_{k(\theta)}^a$ almost surely.
Hence, $\prob{ R_{i,k(\theta)} > 1 \middle| X\sim f_\theta^a } > 0$.
By decomposing $\{  R_{i,k(\theta)} > 1 \} = \bigcup_{n=1}^{\infty}\{ R_{i,k(\theta)} \geq 1 + 1/n \}$, there exists some $n_0$ such that $\prob{ R_{i,k(\theta)} \geq 1 + 1/n_0 \middle| X\sim f_\theta^a } > 0$ (otherwise, the probability of the right hand size is zero).
By setting $\epsilon = 1/n_0$, we have $\prob{ R_{i,k(\theta)} \geq 1 + \epsilon \ \middle| X\sim f_\theta^a } > 0$.
Accordingly, $\lim_{\eta\to\infty} \Lambda(\eta) = \infty$.

Combining that $\Lambda(0) = 1$, the existence of a neighborhood in which $\Lambda(\eta) < 1$, continuity of $\Lambda(\eta)$, and $\lim_{\eta\to\infty} \Lambda(\eta) = \infty$, there exists some $0 < \eta_i^*$ such that $\Lambda(\eta_i^*) = 1$.
Uniqueness of $\eta_i^*$ stems from the convexity of $\Lambda$ in $\eta$, i.e., since
\begin{align*}
    \partialDeriveSecond{\eta}\Lambda = \expValDist{ (R_{i,k(\theta)})^{\eta} (\ln R_{i,k(\theta)})^2 }{f_\theta^a}\geq0
    .
\end{align*}

Applying the Chernoff Bound yields that
\begin{align*}
    \prob{ \text{select $i$} | \theta, n }
    &=
    \prob{ L_{i, k(\theta)}\geq \gamma | \theta, n}
    \\
    &\leq
    2^{-\eta\gamma} \expValDist{2^{\eta L_{i,k(\theta)}}}{f_\theta^a}
    \\
    &=
    2^{-\eta\gamma}\times \left(\expValDist{ \left( \frac{f_{i}^a(X)}{f_{k(\theta)}^a(X)} \right)^{\eta} }{ f_\theta^{a} }  \right)^{\tau}
    \\
    &=
    2^{-\eta\gamma}\times\Lambda(\eta)
\end{align*}
for any $\eta > 0$.
By setting $\eta = \eta_i^*$, we obtain that $\prob{ \text{select $i$} | \theta, n } \leq 2^{-\eta_i^*\gamma}$.


\subsection{Proof of Lemma \ref{lemma: exp family clustering properties}}
\label{lemma: exp family clustering properties proof}
Recall that, for each action $a$, there are functions $\zeta_a$, $T_a$, and $A_a$ such that $f_h^a$ can be decomposed as $f_h^a = \zeta_a(x)\exp\left\{ \eta_h T_a(x) - A_a(\eta_h) \right\}$.
To ease tractability, we drop the action subscript $a$ from these functions.

\subsubsection{Proof of 1)}
\textit{Proof sketch}: We first prove that the \TVDtxt between two members of the same \SEF is “V”- or “U”-shaped in the distribution parameter when the other parameter is fixed. Then, by noticing that $\equivClusterIAEps{\mathcal{U}}$ can induce a partition of the interval $[\min\{\eta_i,\eta_j\}, \max\{\eta_i,\eta_j\}]$, we can leverage the previous claim to show that there is always some $i'\in \equivClusterIAEps{\mathcal{U}}$ whose \TVDtxt from $f_l^a$ does not exceed $\varepsilon_a$ to conclude that $l\in\equivClusterIAEps{\mathcal{U}}$.

We start by showing that \TVDtxt is monotone in the distribution parameters when one distribution is fixed:
\begin{proposition}
    \label{proposition: TVD is monotone in the parameter}
    Let $g(x; \eta) = \zeta(x)\exp\left\{ \eta T(x) - A(\eta) \right\}$.
    If $T(x)$ is monotone in $x$, then $\TVD{f_i^a - g}$ is non-decreasing in $\eta$ when $\eta > \eta_i$ and non-increasing in $\eta$ when $\eta < \eta_i$.
\end{proposition}
\begin{IEEEproof}
    Denote the \CDF of $g(x; \eta)$ as $G(x; \eta) \triangleq \int_{-\infty}^x g(t; \eta) dt$.
    Assume $\eta > \eta_i$ and compute the \LLR between $g(x; \eta)$ and $f_i^a(x)$:
    \begin{align*}
        \log \frac{g(x; \eta)}{f_i^a(x)}
        &=
        \log \exp\left\{
            (\eta - \eta_i) T(x) - (A(\eta) - A(\eta_i))
        \right\}
        \\
        &=
        [(\eta - \eta_i) T(x) - (A(\eta) - A(\eta_i))]\log e
        .
    \end{align*}
    That is, $\log g(x; \eta)/f_i^a(x)$ is linear in $T(x)$, and in particular, monotone in $x$.
    Note that since $\eta > \eta_i$, the trends of $T(x)$ are preserved.
    Assume $T$ is increasing in $x$.
    
    Define $\mathcal{R}_g \triangleq \{x: g(x; \eta) \geq f_i^a(x) \} = \{ x: T(x) \geq \tau_{\eta_i,\eta} \} = \{ x: x \geq T^{-1}(\tau_{\eta_i,\eta}) \}$, where $\tau_{\eta_i,\eta}$ $=$ $(A(\eta) - A(\eta_i))/(\eta - \eta_i)$ and $T^{-1}$ exists since $T$ is monotone.
    Computing the \TVDtxt:
    \begin{align*}
        \TVD{f_i^a - g}
        &=
        \int_{ \mathcal{R}_g } (f_i^a(x) - g(x; \eta)) dx
        \\
        &=
        (1-F_i^a(T^{-1}(\tau_{\eta_i,\eta})) - (1-G(T^{-1}(\tau_{\eta_i,\eta}); \eta))
        \\
        &=
        G(T^{-1}(\tau_{\eta_i,\eta}); \eta) - F_i^a(T^{-1}(\tau_{\eta_i,\eta}))
        \\
        &=
        \sup_{t\in\mathbb{R}}
            G(T^{-1}(t); \eta) - F_i^a(T^{-1}(t))
        \\
        &=
        \sup_{t\in\mathbb{R}}
            \expValDist{\indicator{T(X)\geq t}}{g}
            -
            \expValDist{\indicator{T(X)\geq t}}{f_i^a}
    \end{align*}
    Since $T$ is monotone and $\indicator{y \geq t}$ is non-decreasing in $y$, we can apply \cite[Lemma~3.4.2(i)]{lehmann2006testing}, to obtain that both $\expValDist{\indicator{T(X)\geq t}}{g}$ and $\expValDist{\indicator{T(X)\geq t}}{f_i^a}$ are non-decreasing in their respective parameters, $\eta$ and $\eta_i$ for any $t$.
    Accordingly, since the supremum of non-decreasing functions is non-decreasing, the \TVDtxt is non-decreasing in $\eta > \eta_i$.

    When $T$ is decreasing in $x$, we have $\mathcal{R}_g = \{ x: x \leq T^{-1}(\tau_{\eta_i,\eta}) \}$.
    Thus, the \TVDtxt becomes
    \begin{align*}
        \TVD{f_i^a - g}
        &=
        \sup_{t\in\mathbb{R}}
            F_i^a(T^{-1}(t)) - G(T^{-1}(t); \eta)
    \end{align*}
    Leveraging \cite[Lemma~3.4.2(ii)]{lehmann2006testing}, for any fixed $x_0$ and $\eta\leq \eta'$, $G(x_0;\eta)\geq G(x_0;\eta')$, i.e., $G$ is monotone $\eta$.
    Namely, when $\eta$ increases, $G$ is non-increasing, and the \TVDtxt non-decreasing in $\eta > \eta_i$ as before.
    
    The proof that the \TVDtxt is non-increasing in $\eta$ when $\eta < \eta_i$ is similar by reversing $\eta$ and $\eta_i$, i.e., reversing the roles of $f_i^a(x)$ and $g(x)$.
\end{IEEEproof}

Now that we have shown that \TVDtxt is monotone when moving from a fixed member of the \SEF to either direction, we are ready to show that the clusters are contiguous in the parameter space.
Namely, we want to show that if $f_i^a$ and $f_j^a$ are parametrized by $\eta_i$ and $\eta_j$, respectively, and $i, j\in\equivClusterIAEps{\mathcal{U}}$, then $l\in\equivClusterIAEps{\mathcal{U}}$ for any $l\in\mathcal{H}$ such that $\eta_l\in[\min\{\eta_i,\eta_j\}, \max\{\eta_i,\eta_j\}]$.

Without loss of generality, assume that $\eta_i < \eta_j$.
Since $j\in\equivClusterIAEps{\mathcal{U}}$, there exists some $2\leq\xi_{ij}\in\mathbb{N}$ and a sequence of hypothesis indices $i=l_1, l_2, \dots, l_{\xi_{ij}} = j$ such that $\TVD{f_{l_t}^a - f_{l_{t+1}}^a}\leq \varepsilon_a$ for any $1\leq t\leq \xi_{ij}-1$.

Without loss of generality, assume that $\eta_{l_1} \leq \eta_{l_2}\leq \dots\leq \eta_{l_{\xi_{ij}}}$.
If $\eta_l < \eta_{l_2}$, we have $\TVD{f_i^a-f_l^a} \leq \TVD{f_i^a-f_{l_2}^a}\leq \varepsilon_a$ from Proposition \ref{proposition: TVD is monotone in the parameter}, so $l\in\equivClusterIAEps{\mathcal{U}}$.

Otherwise, $\eta_l > \eta_{l_2}$.
If $\eta_l < \eta_{l_3}$, we have $\TVD{f_{l_2}^a-f_k^a}\leq \TVD{f_{l_2}^a-f_{l_3}^a} \leq \varepsilon_a$ from from Proposition \ref{proposition: TVD is monotone in the parameter} and $l\in\equivClusterIAEps{\mathcal{U}}$ once again.
By induction on $t < \xi_{ij}$, we have $\TVD{f_{l_t}^a-f_l^a}\leq \TVD{f_{l_t}^a-f_{l_{t+1}}^a} \leq \varepsilon_a$ and $l\in\equivClusterIAEps{\mathcal{U}}$.
Finally, if $\eta_{\xi_{ij}-1} < \eta_l$, $\TVD{f_{l_{\xi_{ij}-1}}^a-f_l^a}\leq \TVD{f_{l_{\xi_{ij}-1}}^a-f_{l_{\xi_{ij}}}^a} = \TVD{f_{l_{\xi_{ij}-1}}^a-f_j^a} \leq \varepsilon_a$ and the process to identify which interval $[\eta_{l_t}, \eta_{l_{t+1}}]$ contains $\eta_l$ terminates.
Notably, in all cases $l\in\equivClusterIAEps{\mathcal{U}}$ and the proof of 1) is complete.

\subsubsection{Proof of 2) and 3)}
Before proving 2) and 3), we leverage the inequality $\ln x \leq x-1$, or more precisely $\ln (1/x)\geq 1-x$, to obtain that $\DKLD{\theta}{i,}{k(\theta)}{a}\geq (1-\expValDist{ f_{i}^{a}(X) / f_{k(\theta)}^{a}(X) }{ f_\theta^{a} })/\ln 2$.
Thus, when
\begin{align}
    \label{eq: sufficient condition for clustering}
    \expValDist{ \frac{ f_{i}^{a}(X) }{ f_{k(\theta)}^{a}(X) } }{ f_\theta^{a} } < 1
    ,
\end{align}
then $\DKLD{\theta}{i,}{k(\theta)}{a} > 0$.
Namely, to show 2) and 3), it suffices to show that \eqref{eq: sufficient condition for clustering} holds when $k(\theta)$ are selected as described in Lemma \ref{lemma: exp family clustering properties}.
While this condition may fail in general, we will show shortly that it holds for the \SEF.

To evaluate \eqref{eq: sufficient condition for clustering}, we present the following key proposition:
\begin{proposition}
    \label{prop: exp family sufficient condition}
    If $\{f_h^a\}_{h\in\mathcal{H}}$ are from the \SEF.
    Then, $\ln \expValDist{ g(X; \eta) / f_{k(\theta)}^a(X) }{ f_{\theta}^{a} } = A(\eta-\eta_{k(\theta)}+\eta_{\theta}) - A(\eta_{\theta}) - [A(\eta) - A(\eta_{k(\theta)})]$.
\end{proposition}
\begin{IEEEproof}
    The expectation in Eq. \eqref{eq: sufficient condition for clustering} is evaluated to
    \begin{align*}
        \expValDist{ \frac{ g(X; \eta) }{ f_{k(\theta)}^a(X)}}{ f_{\theta}^{a} }
        &=
        \expValDist{ \frac{ \exp\{\eta T(X) - A(\eta)\} }{ \exp\{\eta_{k(\theta)} T(X) - A(\eta_{k(\theta)})\} }}{ f_{\theta}^{a} }
        \\
        &=
        \frac{e^{-A(\eta)}}{e^{-A(\eta_{k(\theta)})}}
        \expValDist{ e^{ (\eta-\eta_{k(\theta)}) T(X) } }{ f_{\theta}^{a} }
    \end{align*}
    Note that the expectation in the last expression is the \MGF of $T(X)$, $\expVal{e^{s T(X)}}$, evaluated at $s = \eta-\eta_{k(\theta)}$.
    Accordingly,
        \begin{align*}
        \expValDist{ \frac{ g(X; \eta) }{ f_{k(\theta)}^a(X)}}{ f_{\theta}^{a} }
        =
        \frac{e^{-A(\eta)}}{e^{-A(\eta_{k(\theta)})}}
        \times
        e^{A(\eta-\eta_{k(\theta)} + \eta_{\theta}) - A(\eta_{\theta})}
        .
    \end{align*}
    Since both sides are positive, taking the natural logarithm on both sides yields the desired result.
\end{IEEEproof}

We now study the logarithm from Proposition \ref{prop: exp family sufficient condition}.
Define:
\begin{align*}
    \Lambda(s)
    \triangleq
    A(s+\eta_{\theta}) - A(\eta_{\theta}) - [A(s+\eta_{k(\theta)}) - A(\eta_{k(\theta)})]
\end{align*}
Notably, $\ln \expValDist{ g(X; \eta) / f_{k(\theta)}^a(X) }{ f_{\theta}^{a} } = \Lambda(\eta-\eta_{k(\theta)})$ and $\Lambda(0) = 0$.
We claim the following:
\begin{proposition}
    \label{prop: Lambda non-increase}
    If $k(\theta) = \argmax{\{ \eta_l \}}{l\in\equivClusterThetaAEps{\mathcal{U}}}$, then $\Lambda(s)$ is non-increasing in $s$.
\end{proposition}
\begin{IEEEproof}
    We derive with respect to $s$:
    \begin{align*}
        \partialDerive{s}\Lambda(s)
        =
        \partialDerive{s}A(s+\eta_{\theta})
        -
        \partialDerive{s}A(s+\eta_{k(\theta)})
    \end{align*}
    Since $A$ is convex, $\partialDeriveSecond{s}A(s)\geq 0$ so  $\partialDerive{s}A(s)$ is non-decreasing in $s$.
    Therefore, when $\eta_{k(\theta)}\geq \eta_{\theta}$, $\partialDerive{s}A(s+\eta_{\theta}) \leq \partialDerive{s}A(s+\eta_{k(\theta)})$, which in turn implies that $\partialDerive{s}\Lambda(s)\leq 0$ so $\Lambda(s)$ is non-increasing.
\end{IEEEproof}
We established that, if $k(\theta) = \argmax{\{ \eta_l \}}{l\in\equivClusterThetaAEps{\mathcal{U}}}$, then for any $\eta > \eta_{k(\theta)}$, $\expValDist{ g(X; \eta) / f_{k(\theta)}^a(X) }{ f_{\theta}^{a} } = e^{\Lambda(\eta-\eta_{k(\theta)})} \leq e^{\Lambda(0)} = 1$.
The expectation in Eq. \eqref{eq: sufficient condition for clustering} is non-increasing in $\eta > \eta_{k(\theta)}$, thereby proving 2).

We now turn to the proof of part 3), for which we require the following counterpart of Proposition \ref{prop: Lambda non-increase}:
\begin{proposition}
    \label{prop: Lambda non-decrease}
    If $k(\theta) = \argmin{\{ \eta_l \}}{l\in\equivClusterThetaAEps{\mathcal{U}}}$, then $\Lambda(s)$ is non-decreasing in $s$.
\end{proposition}
\begin{IEEEproof}
    When $\eta_{k(\theta)} \leq \eta_{\theta}$, $\partialDerive{s}\Lambda(s)\geq 0$ and $\Lambda(s)$ is non-decreasing.
\end{IEEEproof}
This have established that, if $k(\theta) = \argmin{\{ \eta_l \}}{l\in\equivClusterThetaAEps{\mathcal{U}}}$, then for every $\eta < \eta_{k(\theta)}$ (which implies that $\eta - \eta_{k(\theta)} < 0$), we have $\Lambda(\eta - \eta_{k(\theta)})\leq \Lambda(0) = 0$.
As $\eta$ decreases, the difference $\eta - \eta_{k(\theta)}$ decreases which implies that $\Lambda(\eta - \eta_{k(\theta)})$ decreases.
Consequently, its exponent, i.e., Eq. \eqref{eq: sufficient condition for clustering}, decreases as $\eta$ decreases, proving 3).

\subsection{Proof of Lemma \ref{lemma: PHIDELTA pe bound}}
\label{lemma: PHIDELTA pe bound proof}
When $k(\theta)$ that competes against $i$ is selected as in Lemma \ref{lemma: exp family clustering properties}, by applying the Chernoff bound with $\eta = 1$, we have $\prob{ \text{select $i$} | \theta, n } \leq 2^{-\gamma}$ for any $n$.
The rest of the proof is similar to the proof of Lemma \ref{lemma: Phi pe bound}.

\subsection{Proof of Lemma \ref{lemma: IOTA pe bound}}
\label{lemma: IOTA pe bound proof}
As before, pruning $\theta$ after taking $n\in\mathbb{N}$ samples requires that the following is satisfied:
\begin{align}
    \nonumber
    L_{i\theta}(\indexedSet{a}{1}{n}, \indexedSet{x}{1}{n}) \geq \gamma
    &\iff
    \prod_{t=1}^{n}  f_\theta^{a_t}(x_t) \leq 2^{-\gamma}\prod_{t=1}^{n}  f_i^{a_t}(x_t)
\end{align}
Let $\mathcal{R}_{i,\theta}^{n}$ be the decision region for $\hII$ (where $i\neq\theta$) at timestep $n$ in its competition against $\hTheta$, i.e., 
\begin{align}
    \nonumber
    \mathcal{R}_{i,\theta}^{n}
    \triangleq
    \left\{
        \myVec{x}\in\mathbb{R}^{n} : \prod_{t=1}^{n}  f_\theta^{a_t}(x_t) \leq 2^{-\gamma} \prod_{t=1}^{n}  f_{i}^{a_t}(x_t)
    \right\}
\end{align}
We bound the conditional error probability:
\begin{align*}
    \prob{ \hat{\theta} = i | \theta, n }
    &=
    \int_{\mathcal{R}_{i,\theta}^{n}} 
    \prod_{t=1}^{\tau}  f_\theta^{a_t}(x_t)
    d\indexedSet{x}{1}{n}
    \\
    &
    \leq
    2^{-\gamma}
    \int_{\mathcal{R}_{i,\theta}^{n}} 
    \prod_{t=1}^{\tau}  f_i^{a_t}(x_t)
    d\indexedSet{x}{1}{n}
    \\
    &
    \leq
    2^{-\gamma}
    \int_{\mathbb{R}^{n}} 
    \prod_{t=1}^{\tau}  f_i^{a_t}(x_t)
    d\indexedSet{x}{1}{n}
    =
    2^{-\gamma}
\end{align*}
By applying the union bound, we obtain that $p_e(I)\leq (H-1)\times 2^{-\gamma}$ as desired.
\subsection{Proof of Lemma \ref{lemma: per-iteration Na PHIDELTA}}
\label{lemma: per-iteration Na PHIDELTA proof}
We start with the competition between $k(\theta)$ and some $i\in\mathcal{I}$.
By invoking Kolmogorov’s Maximal Inequality \cite[Theorem~22.4]{Billingsley1995_Prob_and_Measure}, we obtain that $\prob{\sup_{t\leq n} \abs{L_{k(\theta), i}} \geq \xi}\leq n\Xi/(\xi^2)$.
Taking the limit $\xi\to\infty$ results in $\prob{\sup_{t\leq n} \abs{L_{k(\theta,\max), i}} = \infty} = 0$, i.e., $\prob{\sup_{t\leq n} \abs{L_{k(\theta,\max), i}} < \infty}=1$.
Recall that $\frac{1}{n}L_{k(\theta), i}\to \DKLD{\theta}{i,}{k(\theta)}{a} > 0$ almost surely, which implies that the Strong Law of Large Numbers in \cite[Theorem~3.1]{Dragalin_etAl_1999_MSPRT_AsympOpt} can be applied, yielding that this competition requires $(1+\littleO{1})\gamma / \DKLD{\theta}{i,}{k(\theta)}{a}$ samples on average.
Since $k(\theta)$ competes against each hypothesis in $\mathcal{I}$, we obtain the result.

\subsection{Proof of Lemma \ref{lemma: N bound general}}
\label{lemma: N bound general proof}
Recall Assumption \ref{assumption: finite LLR variance} that bounds the second moment of the log-likelihood by some $\Xi < \infty$ under any action for any pair of hypotheses.
Assume the last hypothesis pruned is $\hJJ$.
Before we proceed, observe that upon termination, $\Gamma$ has $\abs{ L_{\theta j} } = \gamma + \sigma_{\theta j}(a_{N})$ where $\sigma_{\theta j}(a_{N})\geq 0$ is an overshoot term that depends on the last acquired sample via action $a_N$.
Accordingly, for any event $A$, at the stopping time, we have
\begin{align}
    \nonumber
    \expVal{ \abs{ L_{\theta j} } \times \indicator{ A } | \theta }
    &=
    \expVal{ (\gamma + \sigma_{\theta j}(a_N) ) \indicator{ A } | \theta }
    \\
    \label{eq: LLR mean excess bound helper 1}
    &\leq
    \gamma \prob{A} + \expVal{\sigma_{\theta j} (a_N) | \theta}
    \\
    \label{eq: LLR mean excess bound helper 2}
    &\leq
    \gamma \prob{A} + \frac{\Xi}{ \KLD{f_{\theta}^{a_N}}{f_j^{a_N}} }
    \\
    \label{eq: LLR mean excess bound}
    &=
    \gamma \left(
        \prob{A} + \frac{\Xi}{ \gamma \KLD{f_{\theta}^{a_N}}{f_j^{a_N}} }
    \right)
\end{align}
where Eq. \eqref{eq: LLR mean excess bound helper 1} follows by removing the indicator in the second addend, and Eq. \eqref{eq: LLR mean excess bound helper 2} follows Lorden’s Inequality \cite{lorden1970excess}.
Leveraging Eq. \eqref{eq: LLR mean excess bound}, we bound $\expVal{ L_{\theta j} | \theta}$, starting with the upper bound:
\begin{align*}
    &\expVal{ L_{\theta j} | \theta}
    \leq
    \expVal{ \abs{ L_{\theta j} } | \theta}
    \\
    &=
    \expVal{ \abs{ L_{\theta j} } \times \indicator{\hat{\theta} = \theta} | \theta}
    +
    \expVal{ \abs{ L_{\theta j} } \times \indicator{\hat{\theta} \neq \theta} | \theta}
    \\
    &\leq
    \gamma \left(
        \prob{\hat{\theta} = \theta | \theta} + \prob{\hat{\theta} \neq \theta | \theta} + \frac{2\Xi}{ \gamma \KLD{f_{\theta}^{a_N}}{f_j^{a_N}} }
    \right)
    \\
    &=
    \gamma \left(
        1 + \frac{2\Xi}{ \gamma \KLD{f_{\theta}^{a_N}}{f_j^{a_N}} }
    \right)
\end{align*}
For a lower bound, we use $x \geq -\abs{x}$ and Eq. \eqref{eq: LLR mean excess bound}:
\begin{align*}
    &\expVal{ L_{\theta j} | \theta}
    =
    \expVal{ L_{\theta j} \indicator{\hat{\theta} = \theta} | \theta}
    +
    \expVal{ L_{\theta j} \indicator{\hat{\theta} \neq \theta} | \theta}
    \\
    &
    \geq
    \gamma \prob{\hat{\theta} = \theta | \theta}
    -
    \expVal{ \abs{ L_{\theta j} } \times \indicator{\hat{\theta} \neq \theta} | \theta}
    \\
    &
    \geq
    \gamma \left(
        \prob{\hat{\theta} = \theta | \theta}
        -
        \prob{\hat{\theta} \neq \theta | \theta}
        -
        \frac{\Xi}{ \gamma \KLD{f_{\theta}^{a_N}}{f_j^{a_N}}}
    \right)
    \\
    &
    \geq
    \gamma \left(
        1-2\times 2^{-\gamma}
        -
        \frac{\Xi}{ \gamma \KLD{f_{\theta}^{a_N}}{f_j^{a_N}}}
    \right)
\end{align*}
where the last transition follows since $p_e(\Gamma)\leq 2^{-\gamma}$.
Now, we isolate $\expVal{N|\theta, \Gamma}$:
\begin{align*}
    \expVal{ L_{\theta j} | \theta}
    &=
    \expVal{ \sum_{t=1}^N \log\frac{ f_j^{a_t} (X_t) }{ f_\theta^{a_t} (X_t) } \middle| \theta }
    \\
    &=
    \sum_{a\in\mathcal{A}}
    \expVal{N_a^I|\theta} \KLD{ f_\theta^a }{ f_j^a }
    \\
    &=
    \expVal{N|\theta, \Gamma} \times \sum_{a\in\mathcal{A}} \frac{ \expVal{N_a^I|\theta} }{ \expVal{N|\theta, \Gamma} }  \KLD{ f_\theta^a }{ f_j^a }
    \\
    &=
    \expVal{N|\theta, \Gamma} \times \expValDist{ \KLD{ f_\theta^A }{ f_j^A } | \Gamma }{A\sim\lambda_\theta}
\end{align*}
With the computed bounds on $\expVal{ L_{\theta j} | \theta}$, we have $\expVal{N|\theta, \Gamma} = (1\pm\littleO{1}) \gamma / \min_{i\neq \theta\in\mathcal{H}} \expValDist{ \KLD{ f_\theta^A }{ f_i^A } | \Gamma }{A\sim\lambda_\theta}$.


\subsection{Proof of Theorem \ref{theorem: opt action seq is NP-Hard}}
\label{theorem: opt action seq is NP-Hard proof}
We reduce the multi-iteration \SHT problem to the problem of constructing the \MWDT, which is NP-hard (a proof sketch of NP-Completeness can be found in \cite{garey_johnson2002computers}).
The reduction is similar to the reduction sketched in \cite{Gan_Jia_Li2021_Decision_Tree_SHT}, but some details differ, as we elaborate soon.

In the multi-iteration \SHT, one is tasked with applying actions from a given set $\mathcal{A}$ to recover $\theta$, where each action is associated with the Wald/Armitage Tests.
Formally, denote by $2^{\mathcal{H}}$ the power set of $\mathcal{H}$.
The Wald/Armitage Tests are functions $\mathfrak{T}_{WA}:\mathcal{A}\times2^{\mathcal{H}}\to 2^{\mathcal{H}}$ mapping the given action $a$ and a given set of candidate hypotheses $\mathcal{U}$ to $\mathcal{U}^{\prime}$ such that $\mathcal{U}^{\prime}\subset\mathcal{U}$ and $\theta\in\mathcal{U}^{\prime}$.
When applying action $a\in\mathcal{A}$ to separate the set $\mathcal{U}$, a cost of $\expVal{N_a^\Gamma|\theta, \mathcal{U}}\in[0, \infty)$ is incurred.

The goal in multi-iteration \SHT is to construct an action sequence $a_1, a_2,\dots$ that minimizes $\frac{1}{\abs{\mathcal{H}}}\sum_{\theta}\expVal{N_{a_1}^\Gamma|\theta, \mathcal{H}} + \sum_{r=2}^{R_{\theta}^{\Gamma}} \expVal{N_{a_r}^\Gamma|\theta, \mathcal{U}_r}$ for any $\theta$, where $\mathcal{U}_r$ is the hypotheses candidate set at iteration $r$ (i.e., $\mathcal{U}_r = \mathfrak{T}_{WA}(a_{r-1}, \mathcal{U}_{r-1})$ with $\mathcal{U}_1 = \mathcal{H}$) and $R_{\theta}^{\Gamma}$ is the number of iterations required to correctly identify $\theta$ under $\Gamma$. 

Note that for \emph{causal} \SHT procedures, the first action $a_1$ must be fixed.
Once fixed, the use of $a_1$ induces several sub-\SHT procedures, each of which is required to separate a given set of candidates or terminate (if the candidate set is a singleton).
Although it may change the proof that multi-iteration \SHT is NP-hard, we argue that even the non-causal multi-iteration \SHT that is forced to take samples despite the knowledge of $\theta$ is NP-hard, which in turn implies that the causal multi-iteration \SHT is NP-hard as well.

In the \MWDT settings, one is tasked with identifying an item of interest, $x_0$ from a given set of items $\mathcal{X}$ using a set of tests $\mathcal{T}$, where each test $t\in\mathcal{T}$ is a testing function $t: 2^{\mathcal{X}}\to 2^{\mathcal{X}}$ associated with a weight $w:\mathcal{T}\to [0,\infty)$.
Specifically, a test $t$ maps its input set $\mathcal{Y}\subseteq\mathcal{X}$ to $\mathcal{Y}^{\prime}\subset\mathcal{Y}$ with $x_0\in\mathcal{Y}^{\prime}$ and incurs a cost of $w(t)$.

The goal in \MWDT is to construct a directed tree $T = (2^{\mathcal{X}}, E)$ whose sum of weights over all paths from the root to a leaf (singleton out of $\mathcal{X}$) is minimal.
Edges are constructed using the tests $\mathcal{T}$, i.e., an edge $(u, v)\in E$ if and only if some $t\in\mathcal{T}$ exists such that $t(u) = v\in 2^{\mathcal{X}}$.
In this case, the edge is assigned a weight of $w(t)$.

The reduction goes as follows.
Given an \MWDT instance $(\mathcal{X}, \mathcal{T}, w)$, set $\mathcal{H}\gets\mathcal{X}$.
Construct an invertible mapping $n:\mathcal{T}\to\mathbb{N}$ that assigns each $t\in\mathcal{T}$ a unique natural number from $n(\mathcal{T})\triangleq\intSet{1}{\abs{\mathcal{T}}}$, and set $\mathcal{A}\gets n(\mathcal{T})$.
Finally, for each “action,” assign the expected number of samples to be $\expVal{N_a^\Gamma|x_0, \mathcal{U}}\gets w(n^{-1}(a))$ for any $\mathcal{U}\subseteq\mathcal{X}$.
Now, by applying the optimal multi-iteration \SHT algorithm, $\Gamma^*$, we obtain a procedure that minimizes
\begin{align*}
    &\frac{1}{\abs{\mathcal{H}}}\sum_{\theta=0}^{\abs{\mathcal{H}}-1}\left(
        \expVal{N_{a_1}^\Gamma|\theta, \mathcal{H}} + \sum_{r=2}^{R_{\theta}^{\Gamma^*}} \expVal{N_{a_r}^\Gamma|\theta, \mathcal{U}_r}
    \right)
    \\
    &=
    \frac{1}{\abs{\mathcal{X}}}\sum_{x_0\in\mathcal{X}}\left(
        w(n^{-1}(a_1)) + \sum_{r=2}^{R_{x_0}^{\Gamma^*}} w(n^{-1}(a_r))
    \right)
    \\
    &=
    \frac{1}{\abs{\mathcal{X}}}\sum_{x_0\in\mathcal{X}}\sum_{r=1}^{R_{x_0}^{\Gamma^*}} w(n^{-1}(a_r))
\end{align*}
which can be used to construct the optimal \MWDT (by tracking the actions), since the above total path weight is the minimal one, scaled by a constant factor.
Since \MWDT is NP-hard, multi-iteration \SHT is NP-hard.

\fi
\end{document}